\documentclass[10pt,twocolumn]{article}
\usepackage[margin=2cm]{geometry}
\usepackage[numbers,sort&compress]{natbib}
\usepackage{hyperref}
\usepackage{doi} % makes DOIs clickable
\usepackage[pdftex]{graphicx}
\graphicspath{{../pdf/}{../jpeg/}}
\usepackage{amsmath, amssymb,amsfonts,amsthm}
\theoremstyle{definition}
\newtheorem{lemma}{Lemma}
\newtheorem{theorem}{Theorem}
\newtheorem{definition}{Definition}
\newtheorem{proposition}{Proposition}
\newtheorem{corollary}{Corollary}
\usepackage{algorithmic}
\usepackage{array}
\usepackage{quantikz}
\usepackage{bm}
\usepackage{physics}
\usepackage{tikz}
\usepackage{braket}
\usepackage{xcolor}

\newcommand{\id}{\text{ID}}

\newcommand{\cands}{\text{cands}}

\newcommand{\init}{\text{init}}

\newcommand{\redcircle}{\tikz[baseline=-0.5ex]\draw[red,fill=red,radius=3pt] (0,0) circle;}
\newcommand{\bluecircle}{\tikz[baseline=-0.5ex]\draw[blue,fill=blue,radius=3pt] (0,0) circle;}
\newcommand{\greencircle}{\tikz[baseline=-0.5ex]\draw[green,fill=green,radius=3pt] (0,0) circle;}

\usepackage{graphicx} % Required for inserting images
\usepackage{authblk}
\usepackage{url} 
\usepackage{orcidlink}
\usepackage{rotating}
\newif\ifanon\anonfalse % \anontrue hides author names (blind copy)
\usepackage{tabularx}
\newcolumntype{L}[1]{>{\raggedright\arraybackslash}p{#1}}

\begin{document}

\title{Quantum Voting Protocol for Centralized and Distributed Voting Based on Phase-Flip Counting}
\ifanon
\author{Author names withheld for review}
\else
\author{Ali~Emre~Aydin \orcidlink{0009-0004-3638-3864} }
\author{Ammar~Daskin \orcidlink{0000-0002-1497-5031} \thanks{Corresponding author. e-mail: adaskin25@gmail.com.} }
\affil{
Department of Computer Engineering\\
Istanbul Medeniyet University\\
Istanbul, Turkiye, 34000
}
\fi
\date{}

\maketitle

\begin{abstract}
We introduce a quantum voting protocol that uses superposition and entanglement to enable anonymous voting with trusted voting machines, in a centralized setting with a single machine and in a distributed setting where several machines are linked to a voting center through quantum channels. A voter selects a candidate through a restricted ballot interface; the machine encodes the vote as a phase-flip operation, controlled by the voter's identity register, on an entangled candidate state. Tallying is performed by interference with a reference branch followed by a measurement of the candidate register, which reproduces the vote distribution without iterative classical counting. Because every tallying run consumes the quantum state, the machines store the cast choices classically and replay the voting stage for each run. The protocol uses only basic quantum gates (Hadamard and controlled-Z). We provide analytical examples (4 voters with 2 candidates and 8 voters with 3 candidates) and numerical experiments that simulate ideal conditions, depolarizing, memory, readout, and gate noise, faulty or mistaken ballots, larger instances with confidence intervals, and sampling convergence. The simulations reproduce the theoretical tally probabilities in the ideal case and show how the tally is distorted by noise and by faulty ballots; no error correction or mitigation is applied. Voter anonymity is formalized as the indistinguishability of tallying transcripts under vote permutations, and an identity-dephasing step renders the identity register exactly maximally mixed under the stated trust assumptions. The numbers of protocol runs needed for exact tallying and for winner identification are analyzed explicitly.
\end{abstract}

\textbf{\textit{Quantum voting protocol, quantum survey protocol, quantum algorithms}}

\section{Introduction}
Trustworthy-efficient voting and survey systems are essential for accurately determining public preferences. A fundamental requirement in such systems is voter anonymity. In classical mail-based or centralized machine voting systems, anonymity can be achieved by generating random voter IDs and distributing them to voters without recording the mapping between identities and assigned IDs. This classical approach can be directly extended to quantum computers using superposition states. In the quantum implementation, we can simply employ a state of the form $\ket{\text{ID-Votes}} = \sum_j \ket{\id_j} \ket{0}_{\text{vote}} \ket{0}_{\text{flag}}$, which represents a superposition of voter IDs with initialized empty vote and flag registers, respectively. Then, when a voter with specific $\id_j$ wishes to cast a vote, this ID serves as a control register to apply the chosen voting operation for their selected candidate. The voting process concludes by measuring the vote register to tally results. This model can easily be extended to distributed environments where anonymous IDs ensure vote anonymity. The security properties of this model are similar to the classical protocol: That means trustworthiness is provided as long as the center generating and storing IDs remains trustworthy. 

One of the earliest works on more advanced anonymous quantum voting protocols is proposed by Vaccaro et al. \cite{vaccaro2007quantum}  where entangled states (ballot states) are shared between voters and the tallyman. Votes are recorded by applying local phase shifts to the ballot state. The tally (continuous phase accumulation for tallying) is obtained by a collective measurement (expectation value of a tally operator) after all votes are cast. A similar approach however based on entanglement of qudits is proposed by Hillery et al. \cite{hillery2006towards}.
Entanglement-based protocols  include also \cite{horoshko2011quantum} where  Bell states are utilized for verification and voters can either cast votes using single qubits or perform entanglement checks to detect curious tallymen; the protocol provides unconditional security for both anonymity and prevents multiple voting. Ref.~\cite{wang2016self} employs two types of entangled states ($|\mathcal{X}_n\rangle$ and $|\mathcal{S}_n\rangle$) to enable anyone to compute the tally while maintaining privacy, non-reusability, verifiability, and fairness without requiring a trusted third party.
Ref.~\cite{wang2021quantum} proposed Bell state-based voting with anonymity trace establishing anonymous entanglement between voters and a tallier using high-dimensional Bell states, allowing voters to privately trace and verify their counted votes while ensuring privacy and non-reusability. 
There are also protocols with superdense coding which leverage multi-particle entanglement and single-particle operations \cite{zhang2025superdense}. 
Conjugate coding approaches utilize unknown quantum states, where ballots are hard to forge and provide information-theoretic anonymity under quantum complexity assumptions \cite{okamoto2008quantum}. 
Quantum memory requirements are eliminated by using sequences of non-orthogonal coherent states with phase shifts, making them implementable with current linear optics technology \cite{xu2022quantum}.
A quantum blockchain based method is proposed in Ref.\cite{sun2019simple}.
 
A more recent voting protocol proposed by Centrone et al. \cite{centrone2022quantum} where GHZ entangled states along with anonymous subroutines are used in the protocol. In particular, the proposed e-voting protocol operates without trusted election authorities or simultaneous broadcasting channels. Their method utilizes an untrusted source of multipartite GHZ entanglement, coupled with classical anonymous subroutines, to achieve a publicly verifiable election. A photonic experiment \cite{marcellino2025experimental} is also provided for this protocol very recently. A secure voting protocol schema is drawn in Ref. \cite{mahmoud2025quantum} by using quantum key distribution. Quantum key distribution and designated verifier signatures are used to ensure confidentiality and authenticity while resisting quantum attacks \cite{prajapat2024qsig}. Ref.\cite{liu2021quantum} presented a protocol based on single particle system.  
There are also private information comparison protocols using different entanglement schemes e.g. \cite{yang2009efficient,chen2010efficient} and anonymous conferencing e.g. \cite{webb2024experimental}. We recommend the survey article \cite{li2024survey} which covers many quantum algorithms and protocols related to different layers of internet.

\textbf{Contribution:}
We introduce a phase-flip counting protocol for trusted voting machines (Section~\ref{sec:models}). The entanglement is held by the machines and the voting center, never by individual voters (who have no access to the quantum registers): a voter only selects a candidate through a {restricted ballot interface (a classical front end or a software interface to a cloud or local quantum machine, with no access to the hardware or registers)}, and the machine applies the corresponding controlled phase flip on that voter's identity branch.

In the \emph{centralized} setting, a single machine holds all quantum registers (the voter-identity register, the candidate register, and the ancilla). In the \emph{distributed} setting, several machines (for example, one per polling station) hold the voter-side halves of entangled candidate pairs, and the voting center (Bob, who keeps the tally of votes) holds the other halves. A machine takes a control qubit from the center, applies a phase flip on behalf of the voter to encode the chosen candidate, and returns the control qubit while retaining its entangled partner until tallying begins. The center repeats this process for all voters. Any disturbance of the machine--center entanglement becomes detectable through verification measurements.

Because every tallying run consumes the quantum state, which cannot be cloned, the machines store the cast choices classically and \emph{replay} the voting stage for each run, on one or several machines in parallel. This is the quantum counterpart of a classical direct-recording electronic (DRE) voting machine and carries the same trust assumption: the machines must be correct. The tally derivation, the complexity analysis, and the anonymity results of this paper are stated for the centralized setting; the distributed setting extends it to machines linked by quantum channels. A variant in which individual voters hold qubits would remove the trust in the machines, but it requires the voters' participation in every run and opens voter-side attacks; we discuss it as future work (Section~\ref{sec:discussion}).

Note that our protocol differs from Vaccaro et al.\cite{vaccaro2007quantum} in several key aspects. While they use continuous phase shifts to accumulate the sum of votes and then perform a phase estimation to extract the tally, our protocol uses discrete phase flips (implemented by controlled-Z gates) on entangled candidate states and then extracts the tally by measuring the candidate register directly in the computational basis. Our method is particularly suited for voting scenarios with multiple candidates and allows for a direct count of votes per candidate without the need for phase estimation.

In addition, in contrast to multi-particle complex constructions, our protocol leverages a minimal gate set-Hadamard and controlled Z gates-to encode votes purely via phase flips on entangled candidate registers. This phase-flip counting approach avoids complex multi-party measurements or large-scale entangled state distribution. Therefore, it can be considered more amenable to near-term hardware. In addition, it offers a direct measurement-based tally that scales linearly in the number of voters and candidates.

{Furthermore, we provide an anonymity analysis based on the exact structure of the identity-register density matrix, including a lightweight identity-dephasing augmentation under which the identity register becomes exactly maximally mixed (under the trust assumptions of Section~\ref{sec:models}), {completeness and soundness bounds for the CHSH-based check of the machine--center entanglement against non-i.i.d.\ and adaptive attacks, a unified adversarial model with a property-by-property security table}, and a structured comparison with existing quantum voting protocols (Table~\ref{tab:comparison}).}

\begin{table*}[htbp]
\centering
\caption{Comparison of the proposed protocol with representative quantum voting protocols.}
\label{tab:comparison}
\small
\resizebox{0.8\textwidth}{!}{%
\begin{tabularx}{\textwidth}{|>{\hsize=0.80\hsize\raggedright\arraybackslash}X|>{\hsize=1.00\hsize\raggedright\arraybackslash}X|>{\hsize=0.90\hsize\raggedright\arraybackslash}X|>{\hsize=1.30\hsize\raggedright\arraybackslash}X|>{\hsize=1.00\hsize\raggedright\arraybackslash}X|>{\hsize=1.00\hsize\raggedright\arraybackslash}X|>{\hsize=1.00\hsize\raggedright\arraybackslash}X|}
\hline
\textbf{Protocol} & \textbf{Quantum resource} & \textbf{Vote encoding} & \textbf{Tally extraction} & \textbf{Trusted parties} & \textbf{Verification} & \textbf{Hardware requirements} \\
\hline
Vaccaro et al.~\cite{vaccaro2007quantum} & Multi-party ballot states & Continuous local phase shifts & Collective measurement of a tally operator (phase estimation) & Tallyman & --- & Multiparty entangled states; high-precision collective measurement \\
\hline
Hillery et al.~\cite{hillery2006towards} & Entangled qudits & Qudit operations & Collective measurement & Election authority & --- & $d$-level systems \\
\hline
Horoshko \& Kilin~\cite{horoshko2011quantum} & Bell states & Single-qubit operations & Direct measurement & Partially trusted tallyman & Bell-state checks against curious tallyman & Qubit Bell pairs \\
\hline
Wang et al.~\cite{wang2016self} & $\ket{\mathcal{X}_n}$, $\ket{\mathcal{S}_n}$ states & Local operations & Self-tallying (anyone can compute) & None & Built-in (verifiability, fairness) & Complex multi-particle entangled states \\
\hline
Wang et al.~\cite{wang2021quantum} & High-dimensional Bell states & Local operations & Tallier with anonymity trace & Tallier & Private vote tracing by voters & Qudit Bell pairs \\
\hline
Centrone et al.~\cite{centrone2022quantum} & GHZ states (untrusted source) & Classical anonymous subroutines & Public tally & None & Publicly verifiable & $N$-party GHZ distribution \\
\hline
\textbf{This work} & Bipartite Bell / generalized pairs (Eq.~\eqref{eq:general_cand}) & Discrete phase flips (controlled-$Z$) & Direct computational-basis measurement via interference (Hadamard test) & {Trusted machine(s) and Center; {voters access the machine only through a restricted ballot interface}} & {CHSH check of machine--Center entanglement (completeness and soundness for tested pairs)} & Hadamard + controlled-$Z$ only; shallow circuits, near-term friendly \\

\hline
\end{tabularx}}
\end{table*}

\paragraph{Organization}
{The remainder of this paper is organized as follows. Section~\ref{sec:models} states the machine model and its trust assumptions. Section~\ref{sec:center} presents the centralized single-machine voting model, establishing the core concepts. Section~\ref{sec:dist} extends this framework to distributed machines linked to the center by quantum channels, with entanglement-based verification. The tallying (counting) method is described in Section~\ref{sec:tallying}. Section~\ref{sec:examples} provides analytical examples for 4 voters with 2 candidates and 8 voters with 3 candidates. Section~\ref{sec:complexity} analyzes the computational and measurement complexity of the protocols. Section~\ref{sec:security} presents the formal security analysis, establishing the identity-dephasing framework and the CHSH verification subroutine. Section~\ref{sec:numerical} presents the numerical simulation results under ideal and noisy conditions and for faulty ballots. Finally, Section~\ref{sec:discussion} discusses practical implementation issues, limitations, and concludes the paper.}

\section{{Models and Trust Assumptions}}
\label{sec:models}

The protocol is analyzed for \emph{trusted voting machines}. {A voter has no access to the quantum hardware or registers: the voter only selects a candidate through a restricted ballot interface (a classical front end or a software interface to a cloud or local quantum machine; assumption M2), and the machine encodes the choice.} Two settings are considered: a \emph{centralized} setting with a single machine, and a \emph{distributed} setting with several machines linked to a voting center by quantum channels. Throughout, $N$ denotes the number of voters and $K$ the number of candidates.

\subsection{{Centralized setting: one trusted machine}}
\label{subsec:machine_model}

\paragraph{Setting.}
A voting machine $\mathcal{M}$ (the Center) holds the identity register, the candidate register, and the ancilla; all entanglement is internal to the machine. A voter $V_j$ selects a ballot $k_j \in \{0,\dots,K-1\}$ (or abstains) through the machine's interface, and the machine applies $U_{\text{vote}}^{(j,k_j)}$ of Eq.~\eqref{eq:vote_unitary} on her identity branch. This is the quantum counterpart of a classical direct-recording electronic (DRE) voting machine.

\paragraph{Assumptions.}
\begin{enumerate}
\item[\textbf{M1}] \emph{Correct execution.} The machine applies exactly the prescribed operations (the voting unitaries, the identity dephasing of Section~\ref{subsec:anonymity}, and the ancilla Hadamard) and performs only the prescribed measurements, namely the ancilla and the candidate register. The identity register is never measured.
\item[\textbf{M2}] {\emph{Restricted ballot interface.} A voter acts on the system only through a ballot interface provided by the machine. This can be a classical front end, or a software interface (for example, an API or a circuit-level front end) to a quantum machine that is hosted in the cloud or run locally. The interface accepts one ballot from each registered and authenticated voter and translates it into the single permitted operation: $Z_k$ on that voter's identity branch for one valid candidate $k$, or the identity for an empty vote. Voters have no access to the quantum hardware or to the machine's registers, and they cannot submit arbitrary circuits, operations, or measurements. Each voter therefore contributes at most one vote. A voter may select the wrong candidate (a mistaken vote); this is indistinguishable from a legitimate vote. If the machine is hosted by a third party, that party belongs to the trusted base (M1, M4). If an interface accepted arbitrary operations, M2 would fail, and the eligibility and secrecy statements below would not apply. Faulty execution is a property of the machine and falls under M1; its effect on the tally is illustrated in Section~\ref{subsec:dishonest}.}
\item[\textbf{M3}] \emph{Replay.} Every tallying run measures, and thereby destroys, the quantum state, and the state cannot be cloned. The machine therefore stores the cast choices as classical records and, for each run, re-prepares the state and replays the entire voting stage from these records. The runs are independent and identically distributed, which justifies the i.i.d.\ samples assumed in Section~\ref{subsec:measurement_complexity}, and they may be distributed over $P$ identical machines. Each run costs $O(Nn)$ gates (Section~\ref{subsec:gate_complexity}), so the total gate cost is $O(R_{\text{total}}\,N n)$, while the wall-clock time is reduced by up to a factor $P$.
\item[\textbf{M4}] \emph{Trusted base.} The machine's classical memory holds the records $(\id_j, k_j)$ needed for replay, or an equivalent sealed form of them. Ballot secrecy against the machine's \emph{internal} memory is therefore not claimed, exactly as for a classical DRE machine whose internal memory is trusted. The anonymity results of Section~\ref{subsec:anonymity} concern everything the machine \emph{exposes}: the tally transcript and, with identity dephasing, the quantum state of the identity register. The random phases of the dephasing step must be generated inside the machine and discarded without being recorded; this is an implementation requirement of the same kind as not logging the order of ballots in a classical machine.
\end{enumerate}

\subsection{Distributed setting: several trusted machines}
\label{subsec:distributed_model}

\paragraph{Setting.}
$P$ machines $\mathcal{M}_1,\dots,\mathcal{M}_P$, each serving the voters of one location, are linked to the voting Center by quantum channels. Each machine holds the machine-side halves of the entangled candidate pairs of its voters; the Center holds the other halves and the control qubits (Section~\ref{sec:dist}). Voters interact only classically with their machine.

\paragraph{Assumptions.}
\begin{enumerate}
\item[\textbf{D1}] \emph{Trusted machines.} Assumptions M1, M2, and M4 apply to every machine. The Center is semi-honest in the sense of Section~\ref{sec:security}.
\item[\textbf{D2}] \emph{Channels and memory.} The quantum channels between the machines and the Center are authenticated, and the machines keep their qubits coherent from distribution until tallying. The machines are institutional hardware, not voters' devices.
\item[\textbf{D3}] \emph{Replay.} As in M3, every tallying run replays the stored ballots. Since the state cannot be cloned, each run consumes a fresh set of entangled pairs, so the entanglement-distribution cost grows with the number of runs.
\item[\textbf{D4}] \emph{Global state.} The tally derivation of Section~\ref{sec:tallying} is formulated for a single global superposition over voter identities. For independent machine--Center pairs we do not claim that the collection of pairs realizes this state. The tally and complexity results of Sections~\ref{sec:tallying}--\ref{sec:complexity} are therefore claimed for the centralized setting, and a rigorous derivation for the distributed setting is left as future work.
\item[\textbf{D5}] \emph{Verification.} The CHSH check of Section~\ref{subsec:verification} tests dedicated pairs between the machines and the Center and detects tampering with the channels during distribution. It is a check of the entanglement, not of the ballots. {Its completeness and soundness are established for the tested pairs without an i.i.d.\ assumption (Section~\ref{subsec:finite_stats}); transferring the result from the tested pairs to the voting pairs is an explicit assumption.}
\end{enumerate}

\subsection{Summary of the two settings}
\label{subsec:model_summary}

Table~\ref{tab:models} summarizes the two settings.

\begin{table*}[htbp]
\centering
\caption{{Centralized setting versus distributed setting.}}
\label{tab:models}
{%
\small
\resizebox{0.9\textwidth}{!}{%
\begin{tabularx}{\textwidth}{|>{\hsize=0.70\hsize\raggedright\arraybackslash}X|>{\hsize=1.15\hsize\raggedright\arraybackslash}X|>{\hsize=1.15\hsize\raggedright\arraybackslash}X|}
\hline
 & \textbf{Centralized (one machine)} & \textbf{Distributed (several machines)} \\
\hline
Quantum systems held by & The machine only (entanglement is internal) & The machines and the Center (entanglement across the quantum channels) \\
\hline
Voter input & {Ballot through the restricted interface (one valid candidate or empty)} & {Ballot through the restricted interface of the machine serving the voter} \\
\hline
Replay of the voting stage & From stored classical records; parallelizable over machines & From stored records at each machine; each run needs fresh entangled pairs \\
\hline
Trusted parties & The machine: correct execution, no recording of dephasing phases, trusted internal memory & The machines (as on the left); the Center for the tally; authenticated channels; reliable quantum memory at the machines \\
\hline
Mistaken or faulty ballots & A mistaken vote is a valid vote; faulty execution distorts the tally (Section~\ref{subsec:dishonest}) & Same; in addition, channel tampering is detected by the CHSH check \\
\hline
Results stated for this setting & Sections~\ref{sec:tallying}, \ref{sec:complexity}, \ref{subsec:anonymity}, and the simulations of Section~\ref{sec:numerical} & Sections~\ref{sec:dist} and \ref{subsec:verification}; the tally results are not re-derived (D4) \\
\hline
\end{tabularx}}}
\end{table*}

\section{Centralized Voting on a Single Machine}
\label{sec:center}
{This section describes the centralized setting of Section~\ref{subsec:machine_model}: all quantum registers reside in a trusted machine, voters only select a candidate {through the restricted ballot interface (M2)}, and the operators below are applied by the machine on behalf of the voters.}

We formulate our centralized voting system using the following notation: Let $N$ denote the number of voters, with $2^n \geq N$ for qubit-based representation, where $n$ represents the number of qubits allocated for voter identification. The parameter $K$ indicates the number of candidates, and $m = \lceil \log_2 K \rceil$ specifies the number of qubits required for candidate encoding.

\subsection{ Encoding of Voter Identities}
Each voter $V_j$ is assigned a unique basis state $\ket{\id_j}$, where $j \in \{0, 1, \dots, N-1\}$. The voter identity space is represented using $n = \lceil \log_2 N \rceil$ qubits, providing sufficient computational basis states to represent all voters when $N \leq 2^n$. The identity register for the complete system is expressed as:
\begin{equation}
\ket{\text{IDs}} = \frac{1}{\sqrt{N}} \sum_{j=0}^{N-1} \ket{\id_j}.
\end{equation}
Alternatively, randomized voter IDs may be employed instead of sequential superposition states for enhanced anonymity.

\subsection{Preparation of Candidate State }
For the candidate representation, we employ entangled states where phase flips encode voting preferences. In the case of two candidates, for instance we can utilize the Bell state (normalization constant omitted for clarity):
\begin{equation}
    \ket{\psi_\cands} = \ket{00} + \ket{11}.
\end{equation}
Voting actions correspond to specific phase modifications on this Bell state, which can be described as:
\begin{equation}
\begin{array}{lcl}
\text{Vote for candidate-0:} & -\ket{00} + \ket{11} & \\
\text{Vote for candidate-1:} & \ket{00} - \ket{11} & \\
\text{Empty vote:} & \ket{00} + \ket{11} & \text{(no phase change)}
\end{array} 
\end{equation}
For systems with more than two candidates, we extend this approach using appropriate entangled states such as W states or generalized entanglement patterns. In general, we can use the following  entangled state on $2m$ qubits:
\begin{equation}
\ket{\psi_\cands} = \frac{1}{\sqrt{K}} \sum_{k=0}^{K-1} \ket{c_k}_A \ket{c_k}_B,
\label{eq:general_cand}
\end{equation}
where $\ket{c_k}$ is the $m$-qubit binary encoding of candidate $k$, and subscripts $A$, $B$ label the two halves of the entangled pairs. For $K = 2$ this reduces to the Bell state.

\subsection{Voting Protocol}

The complete initial quantum state for the centralized voting system is:
\begin{equation}
   \ket{\psi_{\init}} = \frac{1}{\sqrt{N}} \sum_j \ket{\id_j} \ket{\psi_\cands}.
\end{equation}

{For each voter $V_j$ with identity $\ket{\id_j}$ and candidate choice $C_k$, the machine applies} a controlled phase-flip operation to the candidate state. The voting operation is implemented using controlled-Z gates conditioned on the voter's identity register:
\begin{equation}
U_{\text{vote}}^{(j,k)} = \ket{\id_j}\bra{\id_j} \otimes \text{CZ}_k,
\label{eq:vote_unitary}
\end{equation}
Here, the $\ket{\id_j}$ state serves as a control operation to differentiate between voters, ensuring that each voter's operation only affects their designated component of the superposition.   $\text{CZ}_k$ applies the appropriate phase flip to encode vote for candidate $k$. Note that for $\ket{0...0}$ state one can use a phase gate where the first element is negative instead of the standard Pauli-Z.
In more formal way, a vote for candidate $k$ can be encoded by applying a \emph{selective phase-flip operator} $Z_k$ that negates only the $k$-th component of the candidate state while leaving all other components unchanged:
\begin{equation}
Z_k \ket{\psi_\cands} = \frac{1}{\sqrt{K}} \sum_{\ell=0}^{K-1} (-1)^{\delta_{k\ell}} \ket{c_\ell}_A \ket{c_\ell}_B,
\label{eq:selective_phase_flip}
\end{equation}
where $\delta_{k\ell}$ is the Kronecker delta. In other words, $Z_k$ introduces a relative phase of $-1$ on the basis state $\ket{c_k}\ket{c_k}$ with respect to all other basis states. For the two-candidate case ($K = 2$), the selective operators result in the followings:
\begin{equation}
\begin{aligned}
Z_0 \ket{\psi_\cands} &= \frac{1}{\sqrt{2}}\bigl(-\ket{00} + \ket{11}\bigr), \\[4pt]
Z_1 \ket{\psi_\cands} &= \frac{1}{\sqrt{2}}\bigl(\ket{00} - \ket{11}\bigr), \\[4pt]
\mathbb{I}\;\ket{\psi_\cands} &= \frac{1}{\sqrt{2}}\bigl(\ket{00} + \ket{11}\bigr) \quad \text{(empty vote)}.
\end{aligned}
\label{eq:vote_encoding_K2}
\end{equation}

As an additional anonymity safeguard, the Center applies a random identity-dephasing operation at the end of the voting stage and before tallying; this operation commutes with all voting unitaries and leaves the tally statistics unchanged, while rendering the identity register exactly maximally mixed. It is described in detail in Section~\ref{subsec:anonymity}.

\subsection{Control Register Integration}
In quantum computing, a global or relative phase cannot be observed directly in a computational basis measurement. 
Measuring $\ket{\psi_\cands}$ vs. $Z_k\ket{\psi_\cands}$ in the $\ket{c_k,c_k}$ basis yields identical probability distributions: i.e. the phase information is hidden.  To obtain this phase difference, one can interfere the voted state with the original state using an ancilla qubit and a Hadamard gate (a Hadamard test). This converts the phase difference into a population difference in the ancilla’s basis. Post‑selecting on the ancilla being $\ket{1}$ isolates the part of the state where the phase flip actually occurred.

Therefore, to enable efficient tallying as described in Section~\ref{sec:tallying}, we incorporate an ancilla qubit initialized in the Hadamard basis. 
This creates a coherent superposition of the candidate state on two branches:
\begin{equation}
\ket{\psi_{\text{ctrl}}} = \frac{1}{\sqrt{2}}\bigl(\ket{0}\ket{\psi_\cands} + \ket{1}\ket{\psi_\cands}\bigr)=
\frac{1}{\sqrt{2}}\begin{pmatrix}
\ket{\psi_\cands} \\
\ket{\psi_\cands}
\end{pmatrix}.
\end{equation}
The voting operation $Z_k$ is applied \emph{only} to the $\ket{0}$-branch of the ancilla, while the $\ket{1}$-branch retains the original state as a reference. After voting by voter $j$, the joint state becomes:
\begin{equation}
\frac{1}{\sqrt{2}}\bigl(\ket{0}\,Z_k\ket{\psi_\cands} + \ket{1}\ket{\psi_\cands}\bigr).
\end{equation}
A subsequent Hadamard gate on the ancilla transforms this into:
\begin{equation}
\frac{1}{2}\Bigl[\ket{0}\bigl(Z_k\ket{\psi_\cands} + \ket{\psi_\cands}\bigr) + \ket{1}\bigl(Z_k\ket{\psi_\cands} - \ket{\psi_\cands}\bigr)\Bigr].
\end{equation}
Measuring the ancilla in outcome $\ket{1}$ post-selects the \emph{difference state} $Z_k\ket{\psi_\cands} - \ket{\psi_\cands}$, which precisely isolates  the voted candidate's basis state: 
\begin{equation}
Z_k\ket{\psi_\cands} - \ket{\psi_\cands} = -\frac{2}{\sqrt{K}}\,\ket{c_k}_A\ket{c_k}_B,
\label{eq:difference_state}
\end{equation} 
For $K = 2$, this gives the following explicit states:
\begin{equation}
\begin{aligned}
Z_0\ket{\psi_\cands} - \ket{\psi_\cands} &= -\sqrt{2}\,\ket{00}, \\
Z_1\ket{\psi_\cands} - \ket{\psi_\cands} &= -\sqrt{2}\,\ket{11}.
\end{aligned}
\end{equation}
Since the difference states for distinct candidates $k \neq k'$ are orthogonal ($\braket{c_k | c_{k'}} = 0$), these states distinguish the two candidates upon measurement of the candidate register without any ambiguity. Here note that because the difference state has norm $\frac{2}{\sqrt{K}}$, the probability of measuring the ancilla in $\ket{1}$ is $1/K$.

This centralized model provides a foundation for understanding the core quantum voting mechanism, which is extended to distributed environments in the next section.

\section{Distributed Voting through Quantum Channels}
\label{sec:dist}
{This section describes the distributed setting of Section~\ref{subsec:distributed_model}: several trusted machines, each serving the voters of one location, are linked to the voting center through quantum channels. {Voters still interact with their machine only through the restricted ballot interface (M2);} the machines, not the voters, hold the quantum systems.}
The distributed setting uses the voting and tallying mechanisms described in Sections~\ref{sec:center} and~\ref{sec:tallying}{, under the trust assumptions of Section~\ref{sec:models},} while adapting them to machines that are physically separated from the center. {Entanglement distribution lets tampering with the channels, or decoherence in the machines' memories, be detected through verification measurements (Section~\ref{subsec:verification}).}

In the distributed setting, each machine and the voting center (Bob) share entangled qubit pairs representing candidates. For a two-candidate system, we can again employ Bell states:

\begin{equation}
\ket{\psi_\cands} = \ket{00} + \ket{11},
\end{equation}

where the first qubit resides with the machine serving the voter and the second qubit remains with the center. The center prepares identical candidate states for each registered voter and sends the corresponding machine-side qubits through quantum channels. In contrast to the centralized setting, where qubit requirements scale logarithmically with the number of voters, the distributed setting requires a number of qubits growing linearly with the number of voters; moreover, every tallying run replays the voting stage and therefore consumes a fresh set of pairs (Section~\ref{sec:models}). Although more qubits are required, the entanglement distribution ensures that tampering with the channels becomes detectable through verification procedures.

\subsection{Gate Decomposition for Distributed Implementation \cite{nielsen2010quantum}} The main protocol takes advantage of the decomposition of the multi controlled gates with only nearest neighbor interactions.
The multi-controlled operations required for distributed voting can be decomposed into elementary one- and two-qubit gates  even when restricted to nearest-neighbor interactions on quantum hardware.
For arbitrary controlled-unitary operations $\text{CCU}$, where $U$ is applied to the target qubit conditioned on both control qubits, we employ the decomposition:
\begin{equation}
\label{eq_decomposition}
    \text{CCU} = (I \otimes I \otimes A^\dagger) \cdot \text{CCX} \cdot (I \otimes I \otimes B) \cdot \text{CCX} \cdot (I \otimes I \otimes C),
\end{equation}
where $A$, $B$, and $C$ are single-qubit unitaries satisfying $A^\dagger B C = I$ and $A^\dagger X B X C = U$.

This decomposition strategy extends to multi-controlled gates of higher qubit counts, enabling practical implementation using only nearest-neighbor one- and two-qubit operations on current quantum hardware architectures.
Here we should also note that, we can use CCZ or the Toffoli gate (CCNOT) interchangeably by using the following equivalence which may be useful in the applications of multi controlled Z gates:
\begin{equation}
\label{eq_ccx2ccz}
\text{CCX} = (I \otimes I \otimes H) \cdot \text{CCZ} \cdot (I \otimes I \otimes H).
\end{equation}

\subsection{Distributed Voting Protocol}

The voting process proceeds sequentially through coordinated interactions between the machines and the center and can be summarized as follows:

\begin{enumerate}
\item \textbf{Initialization}: {The center prepares an identical entangled candidate state $\ket{\psi_\cands}$ for each voter and distributes the machine-side qubits. For instance, in the case of an entangled pair, it sends the first qubit of each pair to the machine serving the respective voter while retaining the second qubit.}

\item \textbf{Control Qubit Distribution}: As seen in the multi-controlled gate decomposition \eqref{eq_decomposition}, the main unitary is applied only to the target qubit. The operations on the control qubits mainly involve CNOTs. Therefore, for each voting round, the center applies gates to the control qubits which define $\ket{\id_j}$ and sends the last control qubit to the {machine serving voter $V_j$}, initialized in the Hadamard basis $\ket{+} = H\ket{0}$.

\item \textbf{Voting Operation}: {According to the ballot selected by the voter, the machine} applies a controlled-phase flip operation controlled by the received control qubit to encode the candidate choice:
\begin{equation}
U_{\text{vote}}^{(k)} = \text{CCZ}(\text{control}, \text{candidate}_k),
\end{equation}
where the specific phase flip pattern corresponds to {the voter's selected} candidate $k$.

\item \textbf{Qubit Return}: At the end of the controlled operation, the {machine} returns the control qubit to the center while retaining {its} original entangled qubit from the candidate pair obtained at the beginning.
\end{enumerate}

This process repeats sequentially for all voters, with the center maintaining coordination to ensure proper ordering. Each voter's identity branch is processed exactly once, so a voter cannot vote more than once; as in a classical voting machine, this is enforced by the machine's ballot handling and is a trust assumption (Section~\ref{sec:models}), not a consequence of the entanglement.

The distributed setting comes with increased resource requirements and technological complexity, particularly in terms of quantum memory at the machines and channel reliability. Measuring or interfering with the entangled pairs during transmission or storage disturbs the entanglement, which the verification of Section~\ref{subsec:verification} can detect; the machines retain their qubits until tallying and can take part in this verification after the vote. The coordinated sequential process can slow down the voting process.

Because tallying consumes the quantum state, every tallying run requires the machines to replay the stored ballots with a fresh set of entangled pairs.

\section{Determining the Winner and Tallying Votes}
\label{sec:tallying}

The tallying process in this protocol treats the voted states as marked states and leverages their differences from the initial state.  For this purpose, the control register includes an ancilla qubit in the Hadamard basis, which generates the initial unmarked candidate state as:
\begin{equation}
    \frac{1}{\sqrt{2}}\bigl(\ket{0}\ket{\psi_\cands} + \ket{1}\ket{\psi_\cands}\bigr).
\end{equation}

After the voting process (the selective phase-flip operation $Z_k$ for voter $j$'s chosen candidate $k$), the quantum state for voter $j$ becomes:
\begin{equation}
    \frac{1}{\sqrt{2}}\bigl(\ket{0}\,Z_k\ket{\psi_\cands} + \ket{1}\ket{\psi_\cands}\bigr),
\end{equation}
where $Z_k\ket{\psi_\cands}$ differs from the original candidate state $\ket{\psi_\cands}$ only by a relative phase flip on the chosen candidate's basis state (cf.\ Eq.~\eqref{eq:selective_phase_flip}).

Applying a Hadamard gate to the ancilla qubit transforms the state for voter $j$ to:
\begin{equation}
\begin{split}
    \ket{\text{vote}_j} = &\frac{1}{2}\ket{0}\bigl(Z_k\ket{\psi_\cands}
    + \ket{\psi_\cands}\bigr) 
    \\ & + \frac{1}{2}\ket{1}\bigl(Z_k\ket{\psi_\cands} - \ket{\psi_\cands}\bigr).
\end{split}
\label{eq:vote_j_state} % REV: label added for reference from the security section
\end{equation}

In the single-machine model, we control the voting operation using the identity register $\ket{\id_j}$ to obtain the final combined state:
\begin{equation}
    \ket{\text{Votes}} = \frac{1}{\sqrt{N}}\sum_{j=0}^{N-1} \ket{\id_j} \ket{\text{vote}_j}.
\label{eq:votes_full} % REV: label added for reference from the security section
\end{equation}

The tallying process begins by measuring the ancilla qubit. This measurement collapses the state to either the sum or difference component. For vote counting, we post-select on the measurement outcome $\ket{1}$, corresponding to the difference component, yielding:
\begin{equation}
    \ket{\text{Votes}_1} = \frac{1}{\eta} \sum_{j=0}^{N-1} \ket{\id_j} \bigl( Z_{k_j}\ket{\psi_\cands} - \ket{\psi_\cands} \bigr),
    \label{eq:votes_diff}
\end{equation}
where  the normalization constant $\eta$ determined to be $\eta = \sqrt{N} \cdot \frac{2}{\sqrt{K}}$ resulting in a properly normalized state and $k_j$ denotes the candidate chosen by voter $j$.
Using the result from Eq.~\eqref{eq:difference_state}, the difference state for each voter isolates exactly the voted candidate:
\begin{equation}
    Z_{k_j}\ket{\psi_\cands} - \ket{\psi_\cands} = -\frac{2}{\sqrt{K}}\,\ket{c_{k_j}}_A\ket{c_{k_j}}_B.
\end{equation}
Substituting into Eq.~\eqref{eq:votes_diff} and simplifying:
\begin{equation}
    \ket{\text{Votes}_1} = \frac{1}{\sqrt{N}} \sum_{j=0}^{N-1} \ket{\id_j}\,\ket{c_{k_j}}_A\ket{c_{k_j}}_B.
    \label{eq:votes_diff_normalized}
\end{equation}

A subsequent measurement of the candidate register in the computational basis yields outcome $\ket{c_k}\ket{c_k}$ with probability:
\begin{equation}
    p_k = \frac{n_k}{N},
    \label{eq:vote_prob}
\end{equation}
where $n_k = |\{j : k_j = k\}|$ is the number of voters who chose candidate $k$. This establishes that the measurement probability distribution is a faithful representation of the true vote distribution.

\section{Voting Protocol Examples}
\label{sec:examples}
\subsection{Example 1: 4 Voters, 2 Candidates (Blue vs.\ Red)}
\label{sec:example1}

Consider $N = 4$ voters with identity register states $\ket{00}, \ket{01}, \ket{10}, \ket{11}$ and assume that we only have $K = 2$ candidates: Blue (candidate 0, encoded as $\ket{00}$) and Red (candidate 1, encoded as $\ket{11}$). The candidate register is prepared in the Bell state
\begin{equation}
\ket{\psi_\cands} = \frac{1}{\sqrt{2}}\bigl(\ket{00} + \ket{11}\bigr).
\end{equation}
The selective phase‑flip operators are defined as in Eq.~\eqref{eq:selective_phase_flip}:
While $Z_0$ flips the phase of the $\ket{00}$ component, $Z_1$ flips the phase of the $\ket{11}$ component.
We assume the voters' choices are as follows:
\begin{itemize}
    \item Voter $\ket{00}$: Blue \bluecircle \; $\rightarrow$ applies $Z_0$.
    \item Voter $\ket{01}$: Red \redcircle \; $\rightarrow$ applies $Z_1$.
    \item Voter $\ket{10}$: Blue \bluecircle \; $\rightarrow$ applies $Z_0$.
    \item Voter $\ket{11}$: Red \redcircle \; $\rightarrow$ applies $Z_1$.
\end{itemize}

The initial state including the ancilla qubit is
\begin{equation}
\begin{split}
&\frac{1}{\sqrt{2}}\bigl(\ket{0}\ket{\psi_\cands} + \ket{1}\ket{\psi_\cands}\bigr)
 \\
 & =  \frac{1}{2}\Bigl[\ket{0}\bigl(\ket{00} + \ket{11}\bigr) + \ket{1}\bigl(\ket{00} + \ket{11}\bigr)\Bigr].
\end{split}
\end{equation}

For voter $\ket{01}$ (Red vote, $Z_1$), the voted candidate state is
\begin{equation}
Z_1\ket{\psi_\cands} = \frac{1}{\sqrt{2}}\bigl(\ket{00} - \ket{11}\bigr).
\end{equation}
Applying the Hadamard gate on the ancilla transforms the joint state for this voter into

\begin{equation}
\begin{split}
\ket{\text{vote}_{01}} 
= & \frac{1}{2}\Bigl[\ket{0}\bigl(Z_1\ket{\psi_\cands} + \ket{\psi_\cands}\bigr)  \\ 
 & {+ \ket{1}}\bigl(Z_1\ket{\psi_\cands} - \ket{\psi_\cands}\bigr)\Bigr]  \\
= &\frac{1}{2\sqrt{2}}\Bigl[ \ket{0}\bigl( (\ket{00} - \ket{11}) + (\ket{00} + \ket{11}) \bigr)  \\
& + \ket{1}\bigl( (\ket{00} - \ket{11}) - (\ket{00} + \ket{11}) \bigr) \Bigr]  \\
=& \frac{1}{2\sqrt{2}}\Bigl[ \ket{0}\bigl(2\ket{00}\bigr) + \ket{1}\bigl(-2\ket{11}\bigr) \Bigr]  \\
=& \frac{1}{\sqrt{2}}\Bigl[ \ket{0}\ket{00} - \ket{1}\ket{11} \Bigr].
\end{split}
\end{equation}
The difference component (the $\ket{1}$ branch) is $-\frac{1}{\sqrt{2}}\ket{11}$, which is exactly proportional to the voted candidate's basis state. For a Blue vote ($Z_0$), an analogous calculation gives a $\ket{1}$ branch proportional to $-\ket{00}$.

After all voting operations, the joint state before ancilla measurement is
\begin{equation}
\begin{split}
\ket{\text{Votes}} = & \frac{1}{\sqrt{4}} \sum_{j=0}^{3} \ket{\id_j} \otimes \frac{1}{2}\Bigl[\ket{0}\bigl(Z_{k_j}\ket{\psi_\cands}  + \ket{\psi_\cands}\bigr) \\ & \qquad\qquad\qquad+ \ket{1}\bigl(Z_{k_j}\ket{\psi_\cands} - \ket{\psi_\cands}\bigr)\Bigr].
\end{split}
\end{equation}
Measuring the ancilla and post‑selecting on outcome $\ket{1}$ collapses the state to the normalized difference component:
\begin{equation}
\ket{\text{Votes}_1} = \frac{1}{\sqrt{4}} \sum_{j=0}^{3} \ket{\id_j} \otimes \frac{Z_{k_j}\ket{\psi_\cands} - \ket{\psi_\cands}}{\|Z_{k_j}\ket{\psi_\cands} - \ket{\psi_\cands}\|}.
\end{equation}
Using Eq.~\eqref{eq:difference_state} with $K=2$, each difference vector has norm $\sqrt{2}$ and is proportional to the voted candidate's basis state. Specifically,
\begin{align}
Z_0\ket{\psi_\cands} - \ket{\psi_\cands} &= -\sqrt{2}\,\ket{00}, \\
Z_1\ket{\psi_\cands} - \ket{\psi_\cands} &= -\sqrt{2}\,\ket{11}.
\end{align}
Thus the normalized post‑selected state is
\begin{equation}
\begin{split}
\ket{\text{Votes}_1} = & \frac{1}{\sqrt{4}} \Bigl(
\ket{00}(-\ket{00}) + \ket{01}(-\ket{11}) \\ &+ \ket{10}(-\ket{00}) + \ket{11}(-\ket{11}) \Bigr).
\end{split}
\end{equation}

A subsequent measurement of the candidate register (ignoring the identity register) yields outcome $\ket{00}$ (Blue) with probability $2/4 = 0.5$ and outcome $\ket{11}$ (Red) with probability $2/4 = 0.5$, faithfully reflecting the true vote distribution.

\begin{figure*}
    \centering
    \includegraphics[width=0.8\linewidth]{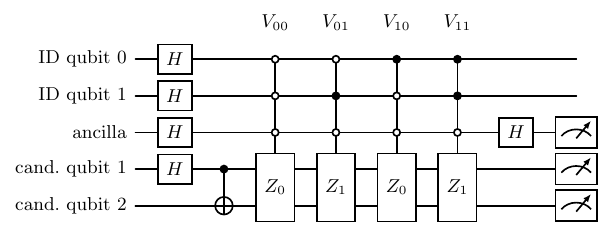}
    \caption{{Voting circuit for Example 1 (4 voters, 2 candidates). The candidate register is prepared in the Bell state, and the identity register and the ancilla in $\ket{+}$. The block $V_{xy}$ applies the phase flip of voter $\ket{xy}$: the identity bits and the ancilla control $Z_0$ (flips $\ket{00}$, Blue) or $Z_1$ (flips $\ket{11}$, Red). Filled circles control on $\ket{1}$, open circles on $\ket{0}$. After the final Hadamard, the ancilla is measured and the run is kept if the outcome is $\ket{1}$; the candidate register is then measured. The identity register is never measured.}}
    \label{fig:circuit-example}
\end{figure*}

{This example is also drawn as a circuit in Fig.~\ref{fig:circuit-example}. Each ballot is one controlled operation: the voter's identity bits and the ancilla control the selective phase flip $Z_k$ on the candidate register (filled circles control on $\ket{1}$, open circles on $\ket{0}$; the open ancilla control applies the flip on the $\ket{0}$ branch). The multi-controlled gates can be decomposed into elementary gates, for example with Eq.~\eqref{eq_ccx2ccz}.}
\subsection{Example 2: 8 Voters, 3 Candidates}
\label{sec:example2}

Consider $N = 8$ voters ($n = 3$ identity qubits) and $K = 3$ candidates represented by the $W$-type state
\begin{equation}
\ket{\psi_\cands} = \frac{1}{\sqrt{3}}\bigl(\ket{00} + \ket{01} + \ket{10}\bigr),
\end{equation}
where the candidate choices are distributed as follows:

\begin{itemize}
    \item Voters $\ket{000}, \ket{011}, \ket{101}$: Candidate 0 \redcircle
    \begin{itemize}
        \item Apply $Z_0$ to flip the phase of the $\ket{00}$ component.
    \end{itemize}
    \item Voters $\ket{001}, \ket{100}, \ket{110}$: Candidate 1 \bluecircle
    \begin{itemize}
        \item Apply $Z_1$ to flip the phase of the $\ket{01}$ component.
    \end{itemize}
    \item Voters $\ket{010}, \ket{111}$: Candidate 2 \greencircle
    \begin{itemize}
        \item Apply $Z_2$ to flip the phase of the $\ket{10}$ component.
    \end{itemize}
\end{itemize}

{\emph{Remark.} In this example we use the compact single-register form of the candidate state, in which candidate $k$ corresponds to the basis state $\ket{c_k}$ of one $m=2$ qubit register, instead of the paired form $\sum_k \ket{c_k}_A\ket{c_k}_B$ of Eq.~\eqref{eq:general_cand}. The derivation uses only the orthonormality of the candidate basis states and is identical; the paired form is needed only when the two halves are held by different parties, as in the distributed setting.}

The initial state including the ancilla qubit is
\begin{equation}
\begin{split}
&\frac{1}{\sqrt{2}}\bigl(\ket{0}\ket{\psi_\cands} + \ket{1}\ket{\psi_\cands}\bigr)\\
&= \frac{1}{\sqrt{6}}\Bigl[\ket{0}\bigl(\ket{00} + \ket{01} + \ket{10}\bigr) \\ & \qquad + \ket{1}\bigl(\ket{00} + \ket{01} + \ket{10}\bigr)\Bigr].
\end{split}
\end{equation}

For voter $\ket{000}$ (Candidate 0), the phase‑flip operator $Z_0$ yields
\begin{equation}
Z_0\ket{\psi_\cands} = \frac{1}{\sqrt{3}}\bigl(-\ket{00} + \ket{01} + \ket{10}\bigr).
\end{equation}
Applying the Hadamard gate on the ancilla transforms the state for this voter into
\begin{align}
\ket{\text{vote}_{000}} 
=& \frac{1}{2}\Bigl[\ket{0}\bigl(Z_0\ket{\psi_\cands} + \ket{\psi_\cands}\bigr) \\ &+ \ket{1}\bigl(Z_0\ket{\psi_\cands} - \ket{\psi_\cands}\bigr)\Bigr] \nonumber \\
=& \frac{1}{2\sqrt{3}}\Bigl[ \ket{0}\bigl( (-\ket{00} + \ket{01} + \ket{10}) 
    \\ & \qquad\qquad+ (\ket{00} + \ket{01} + \ket{10}) \bigr) \nonumber \\
&\qquad + \ket{1}\bigl( (-\ket{00} + \ket{01} + \ket{10}) \\ &\qquad\qquad - (\ket{00} + \ket{01} + \ket{10}) \bigr) \Bigr] \nonumber \\
&= \frac{1}{2\sqrt{3}}\Bigl[ \ket{0}\bigl(2\ket{01} + 2\ket{10}\bigr) \\ & \qquad\qquad+ \ket{1}\bigl(-2\ket{00}\bigr) \Bigr] \nonumber \\
&= \frac{1}{\sqrt{3}}\Bigl[ \ket{0}\bigl(\ket{01} + \ket{10}\bigr) - \ket{1}\ket{00} \Bigr].
\end{align}
The difference component (the $\ket{1}$ branch) is $-\frac{1}{\sqrt{3}}\ket{00}$, which confirms that $Z_0\ket{\psi_\cands} - \ket{\psi_\cands} = -\frac{2}{\sqrt{3}}\ket{00}$ as predicted by Eq.~\eqref{eq:difference_state}.

After the voting operations for all voters, the joint state before ancilla measurement is
\begin{equation}
\begin{split}
\ket{\text{Votes}} = \frac{1}{\sqrt{8}} \sum_{j=0}^{7} \ket{\id_j} \otimes \frac{1}{2}\Bigl[\ket{0}\bigl(Z_{k_j}\ket{\psi_\cands} + \ket{\psi_\cands}\bigr) \\ + \ket{1}\bigl(Z_{k_j}\ket{\psi_\cands} - \ket{\psi_\cands}\bigr)\Bigr].
\end{split}
\end{equation}
Measuring the ancilla and post‑selecting on outcome $\ket{1}$ collapses the state to the (normalized) difference component:
\begin{equation}
\ket{\text{Votes}_1} = \frac{1}{\sqrt{8}} \sum_{j=0}^{7} \ket{\id_j} \otimes \frac{Z_{k_j}\ket{\psi_\cands} - \ket{\psi_\cands}}{\|Z_{k_j}\ket{\psi_\cands} - \ket{\psi_\cands}\|}.
\end{equation}
Using Eq.~\eqref{eq:difference_state}, each difference vector has norm $2/\sqrt{3}$, so the normalized difference state for voter $j$ is simply {$-\ket{c_{k_j}}$}. Thus
\begin{equation}
\ket{\text{Votes}_1} = \frac{1}{\sqrt{8}} \sum_{j=0}^{7} \ket{\id_j} \bigl({-\ket{c_{k_j}}}\bigr),
\end{equation}
which is properly normalized because the $\ket{\id_j}$ are orthonormal and each candidate state has unit norm.

A subsequent measurement of the candidate register (ignoring the identity register) yields outcome {$\ket{c_k}$} with probability
\begin{equation}
p_k = \frac{n_k}{8},
\end{equation}
where $n_0 = 3$, $n_1 = 3$, and $n_2 = 2$. The measurement outcomes and their probabilities are:
\begin{itemize}
    \item $\ket{00}$ (Candidate 0 \redcircle): $3/8 = 0.375$
    \item $\ket{01}$ (Candidate 1 \bluecircle): $3/8 = 0.375$
    \item $\ket{10}$ (Candidate 2 \greencircle): $2/8 = 0.250$
\end{itemize}
These probabilities faithfully reflect the true vote distribution.

\section{Complexity Analysis}
\label{sec:complexity}

The computational cost of the quantum voting protocol comprises two distinct components: the \emph{gate complexity} of implementing the voting operations and the \emph{measurement complexity} (number of repetitions) required to extract the election result with the desired accuracy and confidence. Therefore, we will analyze the complexity in terms of the number of gates and required number of repetitions.

\subsection{Gate Complexity}
\label{subsec:gate_complexity}

For each voter $j$, the central voting system applies the operation $U_{\text{vote}}^{(j,k)}$ defined in Eq.~\eqref{eq:vote_unitary}. This operation is a controlled-$Z_k$ gate, where the control is the identity register $\ket{\id_j}$ and the target is the candidate register. As shown in previous sections, $Z_k$ itself is a multi-controlled $Z$ gate acting on the qubits that encode candidate $k$.

For an $n$-qubit identity register, a multi-controlled $Z$ gate with $n$ controls can be implemented using $O(n)$ elementary quantum gates (e.g., Toffoli and Hadamard gates) with the help of $O(n)$ ancilla qubits, or using $O(n^2)$ gates without ancillas~\cite{nielsen2010quantum}. Since the protocol requires one such operation per voter, the total gate complexity for $N$ voters is
\begin{equation}
    C_{\text{vote}} = O(N \cdot n),
\end{equation}
where the constant factor depends on the specific implementation of the multi-controlled $Z$ gate.

After all votes have been cast, the tallying step involves a single Hadamard gate on the ancilla qubit, followed by a measurement of the ancilla and (upon post‑selection $\ket{1}$ state) a computational basis measurement of the candidate register (a measurement on $\approx \log K$ qubits). Both of these operations require constant time. Thus the overall circuit depth and gate count are dominated by the voting stage and scale linearly with the product of the number of voters and the size of the identity register.

\subsection{Measurement Accuracy and Repetition Overhead}
\label{subsec:measurement_complexity}

The tallying procedure is probabilistic because we must post‑select on the ancilla measurement outcome $\ket{1}$ which is related to number of candidates. The probability of obtaining this outcome in a single run of the protocol can be computed directly from the state before measurement.
After all voters have applied their respective controlled-$Z_k$ operations, the joint state of the identity register, ancilla, and candidate register is
\begin{equation}
\begin{split}
    \ket{\text{Votes}} = \frac{1}{\sqrt{N}} \sum_{j=0}^{N-1} \ket{\id_j} \otimes \frac{1}{\sqrt{2}}\Bigl( \ket{0} Z_{k_j}\ket{\psi_\cands} \\ + \ket{1} \ket{\psi_\cands} \Bigr).
\end{split}
\end{equation}
Applying a Hadamard gate to the ancilla qubit transforms this state into
\begin{equation}
\begin{split}
    & \frac{1}{\sqrt{N}} \sum_{j=0}^{N-1} \ket{\id_j} \otimes \frac{1}{2}\Bigl[ \ket{0}\bigl(Z_{k_j}+I\bigr)\ket{\psi_\cands} \\ & \qquad\qquad\qquad\qquad + \ket{1}\bigl(Z_{k_j}-I\bigr)\ket{\psi_\cands} \Bigr].
\end{split}    
\end{equation}
The probability of measuring the ancilla in the state $\ket{1}$ is the squared norm of the component corresponding to $\ket{1}$:
\begin{align}
    p_{\text{post}} &= \left\| \frac{1}{\sqrt{N}} \sum_{j=0}^{N-1} \ket{\id_j} \otimes \frac{1}{2}\bigl(Z_{k_j}-I\bigr)\ket{\psi_\cands} \right\|^2 \\
    &= \frac{1}{N} \sum_{j=0}^{N-1} \left\| \frac{1}{2}\bigl(Z_{k_j}-I\bigr)\ket{\psi_\cands} \right\|^2,
\end{align}
where we have used the fact that the identity states $\ket{\id_j}$ are orthonormal and that the cross terms vanish. For any candidate $k$, the difference vector is given by Eq.~\eqref{eq:difference_state}:
\begin{equation}
    \frac{1}{2}\bigl(Z_k - I\bigr)\ket{\psi_\cands} = -\frac{1}{\sqrt{K}}\,\ket{c_k}_A\ket{c_k}_B,
\end{equation}
which is a normalized state (since $\|\ket{c_k}_A\ket{c_k}_B\| = 1$) multiplied by the amplitude $-1/\sqrt{K}$. Its squared norm is therefore
\begin{equation}
    \left\| \frac{1}{2}\bigl(Z_k - I\bigr)\ket{\psi_\cands} \right\|^2 = \frac{1}{K}.
\end{equation}
Substituting this into the sum yields
\begin{equation}
    p_{\text{post}} = \frac{1}{N} \sum_{j=0}^{N-1} \frac{1}{K} = \frac{1}{N} \cdot N \cdot \frac{1}{K} = \frac{1}{K}.
    \label{eq:postselect_prob_repeat}
\end{equation}
Notice that this probability depends only on the number of candidates $K$ and is completely independent of both the number of voters $N$ and the actual distribution of votes $\{n_k\}$.

Consequently, to obtain $R$ successful post‑selected samples (i.e., runs where the ancilla reads $\ket{1}$ and a valid vote sample is produced), one must execute the protocol on average
\begin{equation}
    R_{\text{total}} = \frac{R}{p_{\text{post}}} = R \cdot K
\end{equation}
times. The value of $R$ depends on whether the goal is to recover the exact vote counts for all candidates or merely to identify the winner: Here, while the former requires more repetitions than a simple classical counting, the efficiency of the latter is dependent on the vote gap between candidates. Below we analyze these cases separately.

\paragraph{Sampling model.} Each execution is accepted (ancilla reads $\ket{1}$) with probability $1/K$, independently of the votes (Eq.~\eqref{eq:postselect_prob_repeat}). Conditioned on acceptance, the candidate register returns candidate $k$ with probability $p_k=n_k/N$, and different executions are independent. Hence, if the protocol is executed until $R$ acceptances have occurred, the $R$ accepted outcomes are independent and identically distributed (i.i.d.) draws from $p=(p_1,\dots,p_K)$, and the failure probability $\delta$ in the bounds below refers only to these draws. The number of executions $T$ is random with $\mathbb{E}[T]=KR$. Since the number of acceptances among $T$ executions is binomial with success probability $1/K$, the multiplicative Chernoff bound gives $\Pr[T>2KR]\le e^{-R/4}$, because $2KR$ executions contain fewer than $R$ acceptances with at most this probability. Thus $2KR$ executions suffice with probability at least $1-e^{-R/4}$, and this failure probability is below $\delta'$ whenever $R\ge 4\ln(1/\delta')$.

Hoeffding-type bounds give \emph{sufficient} sample sizes. For each task below we therefore also state (i) a variance-sensitive sufficient bound (Bernstein) and (ii) a lower bound valid for any procedure that uses only the accepted outcomes, so that the dependence on $N$, $\Delta$, $\delta$ and $K$ can be read off from matching upper and lower bounds.

\subsubsection{Exact Tallying of All Votes}
\label{subsubsec:exact_tallying}

Suppose we wish to determine the integer vote counts $n_k$ for all $K$ candidates. It suffices that $|\hat p_k-p_k|<\varepsilon=1/(2N)$ for every $k$, because then $N\hat p_k$ rounds to $n_k$. By Hoeffding's inequality~\cite{hoeffding1963}, $\Pr[|\hat p_k-p_k|\ge\varepsilon]\le 2e^{-2R\varepsilon^2}$ for $R$ i.i.d.\ samples. A union bound over the $K$ candidates requires $2Ke^{-2R\varepsilon^2}\le\delta$, i.e.\ $R\ge \frac{1}{2\varepsilon^2}\ln\frac{2K}{\delta}$. With $\varepsilon=1/(2N)$ this gives the sufficient number of accepted outcomes
\begin{equation}
    R_{\text{exact}} = 2N^2\ln\!\left(\frac{2K}{\delta}\right),
    \label{eq:R_exact}
\end{equation}
and an expected number of executions $K R_{\text{exact}} = 2KN^2\ln(2K/\delta)$ (at most twice this with probability $\ge 1-e^{-R_{\text{exact}}/4}$).

\emph{Variance-sensitive bound.} If every vote share satisfies $p_k\le q\le \tfrac12$, Bernstein's inequality (variance $p_k(1-p_k)\le q$, deviations bounded by $1$) gives the sufficient size
\begin{equation}
    R_{\text{exact}}^{\text{Bern}} = \Bigl(8N^2 q+\tfrac{4N}{3}\Bigr)\ln\!\left(\frac{2K}{\delta}\right).
\end{equation}
For $q=\tfrac12$ this exceeds Eq.~\eqref{eq:R_exact} only by the lower-order term $\tfrac{4N}{3}\ln(2K/\delta)$, so Hoeffding is preferable when some candidate is close to a majority; Bernstein is smaller when all shares are small, as in elections with many candidates.

\begin{proposition}[Necessity of order $N^2\log(1/\delta)$]
\label{prop:exact_lower}
Let $N\ge 4$ be even and $\delta<1/4$. Any procedure that uses only $R$ i.i.d.\ accepted outcomes and returns the exact tally with error probability at most $\delta$ on every election must satisfy
\begin{equation}
    R \;\ge\; \frac{N^2-4}{2}\,\ln\!\frac{1}{4\delta}.
\end{equation}
\end{proposition}
\begin{proof}
Consider two elections with two candidates and tallies $(N/2,N/2)$ and $(N/2+1,N/2-1)$. The accepted outcomes are i.i.d.\ Bernoulli$(\tfrac12)$ and Bernoulli$(\tfrac12+\tfrac1N)$, respectively, and the two tallies must be told apart. By the Bretagnolle--Huber inequality, the sum of the two error probabilities of any test is at least $\tfrac12\exp(-R\,D)$, where $D=-\tfrac12\ln(1-4/N^2)\le 2/(N^2-4)$ is the Kullback--Leibler divergence between the two Bernoulli laws. Hence the larger error is at least $\tfrac14 e^{-2R/(N^2-4)}$, and requiring it to be at most $\delta$ gives the claim.
\end{proof}
Together with Eq.~\eqref{eq:R_exact}, this shows that the sample cost of reading out the exact tally from the accepted outcomes is $\Theta(N^2\log(1/\delta))$ up to a constant factor and the $\ln K$ term: the quadratic growth in $N$ is intrinsic to sampling-based readout and not an artifact of the Hoeffding bound. It becomes prohibitive for very large $N$, but is acceptable for small- to medium-scale elections.

\subsubsection{Determining Only the Winner}
\label{subsubsec:winner_only}

In many practical scenarios it suffices to identify the candidate with the most votes. Relabel the candidates so that $p_1>p_2\ge p_3\ge\dots\ge p_K$ and let $\Delta=p_1-p_2>0$ be the margin. The decision rule is to output the candidate with the largest number of accepted outcomes. A wrong output requires that some rival $j\ne 1$ receives at least as many accepted outcomes as candidate~1. Since this involves $K-1$ rivals, the required sample size depends explicitly on the number of candidates.

\begin{proposition}[Winner identification]
\label{prop:winner}
If the number of accepted outcomes satisfies
\begin{equation}
    R \;\ge\; R_{\text{winner}} = \frac{2}{\Delta^2}\,\ln\!\left(\frac{K-1}{\delta}\right),
    \label{eq:R_winner}
\end{equation}
then the rule above outputs candidate $1$ with probability at least $1-\delta$. The same holds under the variance-sensitive condition
\begin{equation}
    R \;\ge\; \frac{2(p_1+p_2)+\tfrac23(1+\Delta)\Delta}{\Delta^2}\,\ln\!\left(\frac{K-1}{\delta}\right).
    \label{eq:R_winner_bern}
\end{equation}
\end{proposition}
\begin{proof}
Fix a rival $j$ and set $Y_i=\mathbf 1[X_i=1]-\mathbf 1[X_i=j]\in\{-1,0,1\}$ for the $i$-th accepted outcome $X_i$. Then $\mathbb{E}[Y_i]=p_1-p_j=\Delta_j\ge\Delta$, and $j$ ties or beats candidate~1 only if $\sum_{i=1}^R Y_i\le 0$. Hoeffding's inequality for variables of range $2$ gives $\Pr[\sum_i Y_i\le 0]\le \exp(-R\Delta_j^2/2)\le\exp(-R\Delta^2/2)$. A union bound over the $K-1$ rivals gives failure probability at most $(K-1)e^{-R\Delta^2/2}\le\delta$ under Eq.~\eqref{eq:R_winner}. For Eq.~\eqref{eq:R_winner_bern}, apply Bernstein's inequality instead, using $\mathrm{Var}(Y_i)\le p_1+p_j\le p_1+p_2$ and $\mathbb{E}[Y_i]-Y_i\le 1+\Delta_j$; the resulting exponent is increasing in $\Delta_j$, so the worst case is $\Delta_j=\Delta$.
\end{proof}
For $K=2$, Eq.~\eqref{eq:R_winner} reduces to the familiar two-candidate expression $\frac{2}{\Delta^2}\ln\frac1\delta$; for $K>2$ the union over rivals adds the factor $\ln(K-1)$. The bound does not depend on $N$. The expected number of protocol executions is
\begin{equation}
    \mathbb{E}\bigl[R_{\text{total}}^{\text{winner}}\bigr] = K\,R_{\text{winner}} = \frac{2K}{\Delta^2}\ln\!\left(\frac{K-1}{\delta}\right),
\end{equation}
where the factor $K$ is the exact post-selection overhead and the $\ln(K-1)$ comes from the union bound. We do not claim that the $\ln(K-1)$ term is necessary. In contrast, the factor $1/\Delta^2$ and the $\ln(1/\delta)$ term cannot be avoided:

\begin{proposition}[Necessity of $\Delta^{-2}\log(1/\delta)$]
\label{prop:winner_lower}
Any procedure that uses only $R$ i.i.d.\ accepted outcomes and identifies the winner with error probability at most $\delta<1/4$ on every election with margin $\Delta$ must satisfy $R\ge \frac{1-\Delta^2}{2\Delta^2}\ln\frac{1}{4\delta}$.
\end{proposition}
\begin{proof}
Compare two-candidate elections with shares $(\tfrac{1+\Delta}{2},\tfrac{1-\Delta}{2})$ and the swapped shares. The Kullback--Leibler divergence is $\Delta\ln\frac{1+\Delta}{1-\Delta}\le \frac{2\Delta^2}{1-\Delta^2}$, and the Bretagnolle--Huber argument of Proposition~\ref{prop:exact_lower} applies.
\end{proof}
When the race is a tie or the margin is extremely small, the required $R$ grows as $1/\Delta^2$ and eventually exceeds $R_{\text{exact}}$; in that case exact tallying is the better choice.

\subsubsection{Adaptive Strategy}
\label{subsubsec:adaptive}

Since $\Delta$ is unknown a priori, the number of samples cannot be fixed in advance, and naively re-testing a fixed-sample bound after every batch would invalidate it. The following checkpoint rule keeps the error probability below $\delta$ at all stopping times.
\begin{enumerate}
    \item Choose an initial size $R_0$ and checkpoints $R_t=R_0\,2^{t-1}$, $t=1,2,\dots$, up to a budget $R_{\max}$ (e.g.\ $R_{\text{exact}}$). At checkpoint $t$ use all accepted outcomes collected so far.
    \item Let $\hat G_{ij}$ be the difference of the empirical frequencies of candidates $i$ and $j$, and let $\hat\imath$ be the empirical leader. Set
    \begin{equation}
        \eta_t=\sqrt{\frac{2}{R_t}\ln\frac{K(K-1)\,t(t+1)}{\delta}}.
    \end{equation}
    \item If $\hat G_{\hat\imath j}>\eta_t$ for every $j\ne\hat\imath$, stop and declare $\hat\imath$ the winner.
    \item Otherwise continue to checkpoint $t+1$. If the budget $R_{\max}$ is reached without stopping, report ``no decision at this margin'' and fall back to exact tallying.
\end{enumerate}
\begin{proposition}[Anytime-valid stopping]
\label{prop:adaptive}
With probability at least $1-\delta$ the rule never declares a wrong winner, and it stops at the first checkpoint with $R_t>\frac{8}{\Delta^2}\ln\frac{K(K-1)t(t+1)}{\delta}$ (if the budget allows), so the number of accepted outcomes used is $O\bigl(\Delta^{-2}(\ln\frac K\delta+\ln\ln\frac1\Delta)\bigr)$.
\end{proposition}
\begin{proof}
For an ordered pair $(i,j)$, Hoeffding's inequality gives $\Pr[\hat G_{ij}-G_{ij}>\eta_t]\le e^{-R_t\eta_t^2/2}=\delta/\bigl(K(K-1)t(t+1)\bigr)$ with $G_{ij}=p_i-p_j$. A union bound over the $K(K-1)$ ordered pairs and over checkpoints, using $\sum_t \frac{1}{t(t+1)}=1$, shows that with probability $\ge1-\delta$ we have $\hat G_{ij}-G_{ij}\le\eta_t$ for all $i,j,t$. On this event, stopping with $\hat G_{\hat\imath j}>\eta_t$ for all $j$ implies $G_{\hat\imath j}>0$ for all $j$, i.e.\ $\hat\imath$ is the true winner. The lower-tail version of the same bound gives $\hat G_{1j}\ge\Delta-\eta_t>\eta_t$ as soon as $\eta_t<\Delta/2$, which is the stated condition on $R_t$.
\end{proof}
The price of not knowing $\Delta$ is a constant factor (about $4$) compared to Eq.~\eqref{eq:R_winner}; the rule keeps the full correctness guarantee of the protocol while adapting the measurement cost to the actual margin.

As a summary, one execution of the protocol requires $O(N \cdot n)$ elementary gates for $N$ voters and $n = \lceil\log_2 N\rceil$ identity qubits. The expected number of executions, including the exact post-selection overhead $K$, is $2KN^2\ln(2K/\delta)$ for exact tallying (Eq.~\eqref{eq:R_exact}), which is $\Theta(KN^2\log(1/\delta))$ up to the $\ln K$ term by Proposition~\ref{prop:exact_lower}, and $\frac{2K}{\Delta^2}\ln\frac{K-1}{\delta}$ for winner determination (Eq.~\eqref{eq:R_winner}), which is independent of $N$ for a fixed margin $\Delta$ and has the necessary $\Delta^{-2}\log(1/\delta)$ scaling (Proposition~\ref{prop:winner_lower}). Since every execution replays the voting stage from the stored ballots (Section~\ref{sec:models}), the total gate count is the number of executions times $O(Nn)$, that is, $O(K N^3 n \log(K/\delta))$ for exact tallying and $O\bigl(\frac{K}{\Delta^2}\, N n \log(K/\delta)\bigr)$ for winner determination. The executions are independent and can be distributed over several machines, which reduces the wall-clock time by up to the number of machines.

The gate count of a single execution is therefore efficient for any election size. The number of executions is practical for exact tallying when $N$ is moderate, and for winner determination even when $N$ is very large, provided the margin of victory is not vanishingly small; in the latter case, however, the total work still grows linearly with $N$ because every execution replays all ballots.

\section{Security Analysis}
\label{sec:security}
Quantum communication protocols such as BB84~\cite{bennett2014quantum,nielsen2010quantum} provide security against man-in-the-middle attacks by using quantum mechanics and the detection of basis-information changes through a classical channel. 
The security of quantum protocols is generally analyzed under certain assumptions.  In our analysis, we will also cover anonymity, verifiability, and resistance to common attack vectors by making the following {assumptions:}

\paragraph{Trusted Machines} The machines (in the centralized setting, the single machine) execute the protocol correctly: they apply the prescribed operations, perform only the prescribed measurements, store the ballots as classical records for replay, and do not record the random phases of the identity-dephasing step (assumptions M1--M4 of Section~\ref{sec:models}). {Voters have no access to the quantum hardware or registers (M2).}
\paragraph{Authenticated Quantum Channels} In the distributed setting we assume that there exist authenticated quantum channels between the Center and each machine. This prevents man-in-the-middle attacks during qubit distribution and can be realized using Quantum Key Distribution (QKD) combined with classical authentication~\cite{bennett2014quantum, mehic2020quantum}. The centralized setting needs no quantum channels.

\paragraph{Semi‑Honest Center} For the anonymity guarantees, we assume the Center (Bob) correctly follows the protocol specification, and in particular performs only the measurements prescribed by the protocol (the ancilla and candidate-register measurements of Section~\ref{sec:tallying}). Otherwise, a fully malicious Center that prepares non‑standard initial states or performs intermediate measurements can potentially break anonymity.
{\paragraph{Reliable Quantum Memory} The machines in the distributed setting must maintain coherence of their entangled qubits from distribution until tallying.}

Below, we provide a formal security analysis of the proposed quantum voting protocol by first deriving the exact structure of the identity-register density matrix and formalizing the anonymity guarantee provided against a semi‑honest Center. We subsequently analyze the protocol's resilience against specific attack vectors and describe an entanglement‑based verification subroutine for the distributed setting. The section concludes with a discussion of limitations and directions for future work.

\subsection{Unified Adversarial Model and Security Goals}
\label{subsec:unified}

All security statements of this paper are made in the single model below. For each property we state whether it is \emph{proved} in the model, \emph{assumed} through a trust assumption of Section~\ref{sec:models}, or \emph{not provided}.

\paragraph{Parties and reference behavior.} The parties are the registered voters $V_0,\dots,V_{N-1}$ {(ballot interface only)}, the voting machine(s), the Center, and an adversary. The reference is an ideal voting functionality $\mathcal F_{\text{vote}}$ that accepts at most one ballot from each registered voter, releases nothing before the polls close, and then releases only the tally (or only the winner).

\paragraph{Adversaries.}
\begin{itemize}
\item[\textbf{A1}] \emph{External quantum adversary} on the quantum channels during entanglement distribution (distributed setting): arbitrary, adaptive, non-i.i.d.\ attacks; the classical channels are authenticated.
\item[\textbf{A2}] \emph{Observer of public outputs with auxiliary information}: sees the transcript and holds any side information, including the ballots of colluding voters.
\item[\textbf{A3}] \emph{Semi-honest Center}: follows the protocol and analyzes everything it legitimately holds.
\item[\textbf{A4}] \emph{Dishonest voter}: has only the {restricted ballot interface (M2)}; may try to vote twice, to vote without being registered, or to vote for another voter.
\item[\textbf{A5}] \emph{Benign faults}: gate, channel, and memory noise, and miscalibrated operations.
\item[\textbf{A6}] \emph{Out of scope}: a malicious machine or Center that deviates from M1--M4, denial of service, and coercion or vote buying.
\end{itemize}

\paragraph{Properties and results.} Table~\ref{tab:security} lists the eight properties, the adversaries and assumptions under which they are considered, and the status of each. We give the statement for the one property whose proof is not already in the preceding sections.

\begin{proposition}[Eligibility and one vote per voter]
\label{prop:eligibility}
Let the registered identities be $\{\id_j\}_{j=0}^{N-1}$, and assume M1 and M2. After any sequence of ballot operations the state has the form $\frac1{\sqrt N}\sum_{j}\ket{\id_j}\otimes\ket{\phi_j}$, where $\ket{\phi_j}$ depends only on the single ballot recorded for voter $j$. Consequently the tally weights $n_k$ count each registered voter at most once and each unregistered identity zero times, and an operation controlled on an identity $\id\notin\{\id_j\}$ acts as the identity on the state.
\end{proposition}
\begin{proof}
Every ballot operation $U^{(j,k)}_{\text{vote}}$ is block diagonal in the identity basis (Eq.~\eqref{eq:vote_unitary}) and acts nontrivially only on the branch $\ket{\id_j}$. By M2 the machine applies at most one such operation per registered voter. A controlled operation whose control is an identity outside the register has no support on the state, because the state has zero amplitude on that identity.
\end{proof}
The registration and the authentication of a voter at the machine are classical and are assumed (M2), as in a classical voting machine; the proposition shows that the quantum encoding adds no way around them.

\begin{table*}[htbp]
\centering
\caption{{Security properties in the unified model. ``Proved'' means proved in this paper under the stated adversary and assumptions; ``Assumed'' means obtained from a trust assumption; ``Not provided'' means the protocol makes no claim.}}
\label{tab:security}
{%
\small
\resizebox{0.9\textwidth}{!}{%
\begin{tabularx}{\textwidth}{|>{\hsize=0.70\hsize\raggedright\arraybackslash}X|>{\hsize=0.70\hsize\raggedright\arraybackslash}X|>{\hsize=0.80\hsize\raggedright\arraybackslash}X|>{\hsize=1.80\hsize\raggedright\arraybackslash}X|}
\hline
\textbf{Property} & \textbf{Adversary / assumptions} & \textbf{Status} & \textbf{Basis and scope} \\
\hline
Correctness & A4; M1--M3, no faults (A5 absent) & Proved (statistical) & The published tally or winner is wrong with probability at most $\delta$ (Eq.~\eqref{eq:R_exact}, Propositions~\ref{prop:winner} and~\ref{prop:adaptive}). With faults (A5) correctness degrades; the effect is quantified numerically in Section~\ref{sec:numerical}, without error correction. \\
\hline
Fairness (no early result) & A2, A3; M1 & Assumed & No tallying run takes place before the polls close; this is a procedure of the machine (M1). {The ballot interface gives voters no measurement access, so no voter can read out a partial result.} \\
\hline
Ballot secrecy & A2, A3; M1, M4 & Proved for views V1 and V2; not claimed for V3 & Corollary~\ref{cor:aux} (transcript with arbitrary auxiliary information) and Proposition~\ref{prop:views}(a) (identity register). The complete quantum state of the machine reveals a random voter's ballot in each run (Proposition~\ref{prop:views}(b),(c)) and is part of the trusted base (M4). \\
\hline
Eligibility and one vote per voter & A4; M1, M2 & Proved given M1--M2 & Proposition~\ref{prop:eligibility}. Voter registration and authentication are classical and assumed. \\
\hline
Individual verifiability & --- & Not provided & A voter cannot check that the recorded ballot equals the ballot cast. The machine stores classical records (M3), which could support a classical receipt or paper trail, but this is not part of the protocol and conflicts with receipt-freeness. \\
\hline
Universal verifiability & --- & Partial & Anyone can check from the public transcript that the declared winner or tally follows from the accepted outcomes with the stated confidence (Propositions~\ref{prop:winner}, \ref{prop:adaptive}). That the transcript was produced from the recorded ballots cannot be verified. \\
\hline
Robustness & A4, A5 & Partial & A mistaken ballot is a valid vote; the statistical failure probability is at most $\delta$; noise distorts the tally and no error correction or mitigation is claimed (Sections~\ref{subsec:dishonest}, \ref{subsec:noise}). Not provided against A6. \\
\hline
Tally integrity & A1 (distributed); M1 & Partial & Channel tampering with the entanglement is detected for the tested pairs (Theorem~\ref{thm:soundness}); transfer to the voting pairs is an explicit assumption. Integrity against the machine or the Center is assumed (M1) and not provided against A6. \\
\hline
\end{tabularx}}}
\end{table*}

\subsection{Information‑Theoretic Anonymity}
\label{subsec:anonymity}
Here, we first assume that a semi‑honest Center executes the protocol honestly but attempts to learn the mapping between voter identities $\ket{\id_j}$ and their candidate choices.
Rather than asserting that the identity register is maximally mixed---which, as we show below, is not true in general for the bare protocol---we proceed in three steps: (i) we derive the \emph{exact} structure of the identity-register density matrix, making the off-diagonal terms explicit; (ii) we introduce a lightweight \emph{identity-dephasing} augmentation of the protocol under which the identity register becomes exactly maximally mixed; and (iii) we formalize the operational anonymity guarantee as the indistinguishability of protocol transcripts under vote permutations.

We adopt the following operational definition of anonymity.

\begin{definition}[Anonymity as vote-permutation indistinguishability]
\label{def:anonymity}
Let $v = (k_0, k_1, \dots, k_{N-1})$ denote a vote assignment, and let $T(v)$ denote the \emph{transcript} of the protocol: the random record of all measurement outcomes prescribed by the protocol (the ancilla and candidate-register outcomes of Section~\ref{sec:tallying}) accumulated over all repetitions. The protocol provides \emph{information-theoretic anonymity} against a semi-honest Center if, for any two vote assignments $v$ and $v'$ with identical tallies, $n_k(v) = n_k(v')$ for all $k$, the transcripts $T(v)$ and $T(v')$ are identically distributed. Equivalently, the mutual information between the transcript and the voter-to-vote assignment, conditioned on the tally, is zero.
\end{definition}

We first characterize the exact overlap structure of the per-voter states, which determines the identity-register density matrix.

\begin{lemma}[Overlap structure of voted states]
\label{lem:overlaps}
Let $\ket{\phi_j} := \ket{\text{vote}_j}$ be the normalized joint state of the ancilla and candidate registers after voter $j$'s vote for candidate $k_j$, as given in Eq.~\eqref{eq:vote_j_state}. Then, for any two voters $j, j'$,
\begin{equation}
    \braket{\phi_{j'} | \phi_j} \;=\; 1 - \frac{2}{K}\bigl(1 - \delta_{k_j k_{j'}}\bigr)
    \;=\;
    \begin{cases}
        1, & k_j = k_{j'},\\[4pt]
        \dfrac{K-2}{K}, & k_j \neq k_{j'}.
    \end{cases}
    \label{eq:overlap_structure}
\end{equation}
In particular, for $K = 2$ the voted states of voters choosing different candidates are orthogonal.
\end{lemma}

\begin{proof}
From Eq.~\eqref{eq:vote_j_state}, $\ket{\phi_j} = \frac{1}{2}\bigl[\ket{0}(Z_{k_j} + I) + \ket{1}(Z_{k_j} - I)\bigr]\ket{\psi_\cands}$. Since the selective phase-flip operators are Hermitian and involutory ($Z_k^\dagger = Z_k$, $Z_k^2 = I$), and using the orthonormality of the ancilla basis,
\begin{align}
    \braket{\phi_{j'} | \phi_j}
    &= \tfrac{1}{4}\bra{\psi_\cands}(Z_{k_{j'}} + I)(Z_{k_j} + I)\ket{\psi_\cands} \nonumber\\
    &\quad + \tfrac{1}{4}\bra{\psi_\cands}(Z_{k_{j'}} - I)(Z_{k_j} - I)\ket{\psi_\cands} \nonumber\\
    &= \tfrac{1}{2}\bigl(1 + \bra{\psi_\cands} Z_{k_{j'}} Z_{k_j} \ket{\psi_\cands}\bigr).
\end{align}
Since $Z_{k'} Z_k$ acts diagonally on $\ket{\psi_\cands}$ (Eq.~\eqref{eq:general_cand}) with eigenvalue $(-1)^{\delta_{k\ell} + \delta_{k'\ell}}$ on the component $\ket{c_\ell}_A\ket{c_\ell}_B$, we obtain
\begin{equation}
    \bra{\psi_\cands} Z_{k'} Z_k \ket{\psi_\cands}
    = \frac{1}{K}\sum_{\ell=0}^{K-1} (-1)^{\delta_{k\ell} + \delta_{k'\ell}}
    = 1 - \frac{4}{K}\bigl(1 - \delta_{kk'}\bigr),
\end{equation}
because exactly two terms in the sum equal $-1$ when $k \neq k'$ and none when $k = k'$. Substituting yields Eq.~\eqref{eq:overlap_structure}.
\end{proof}

\begin{theorem}[Structure of the identity register]
\label{thm:anonymity}
After all $N$ voters have applied their voting unitaries and before any measurement, the reduced density matrix of the identity register is
\begin{equation}
    \rho_{ID} = \frac{1}{N} \sum_{j,j'=0}^{N-1} G_{jj'}\, \ket{\id_j}\!\bra{\id_{j'}},
    \qquad
    G_{jj'} = \braket{\phi_{j'}|\phi_j},
    \label{eq:rho_id_exact}
\end{equation}
with $G_{jj'}$ given by Lemma~\ref{lem:overlaps}. This matrix has the following properties:
\begin{enumerate}
    \item \textbf{Uniform diagonal:} $\bra{\id_j}\rho_{ID}\ket{\id_j} = 1/N$ for all $j$; hence any computational-basis measurement of the identity register yields uniform statistics and reveals no vote information.
    \item \textbf{Partition invariance of the off-diagonals:} $G_{jj'}$ depends only on whether $k_j = k_{j'}$, i.e., only on the \emph{partition} of voters into equal-vote classes, and is invariant under any relabeling of the candidates. Consequently, even complete tomography of $\rho_{ID}$ reveals at most which voters voted identically---never which candidate any voter chose.
    \item \textbf{Non-maximal mixedness in general:} unless all overlaps vanish, $\rho_{ID} \neq \mathbb{I}/N$. In particular, for $K \geq 3$ all off-diagonal entries between voters with different choices equal $\frac{K-2}{NK} \neq 0$, and for $K = 2$ the off-diagonal entries within each equal-vote class equal $1/N$.
\end{enumerate}
\end{theorem}

\begin{proof}
The joint state of the identity register, ancilla, and candidate register before any measurement is $\ket{\Psi} = \frac{1}{\sqrt{N}} \sum_{j} \ket{\id_j} \otimes \ket{\phi_j}$ (Eq.~\eqref{eq:votes_full}). Tracing out the ancilla and candidate registers gives $\rho_{ID} = \frac{1}{N}\sum_{j,j'} \braket{\phi_{j'}|\phi_j}\, \ket{\id_j}\!\bra{\id_{j'}}$, which is Eq.~\eqref{eq:rho_id_exact}. Property 1 follows because each $\ket{\phi_j}$ is normalized, $G_{jj} = 1$. Property 2 follows directly from Lemma~\ref{lem:overlaps}: the value of $G_{jj'}$ is a function of the indicator $\delta_{k_j k_{j'}}$ alone; relabeling candidates permutes the values $k_j$ but preserves all equalities $k_j = k_{j'}$, leaving $G$ (and hence $\rho_{ID}$) unchanged. Property 3 is immediate from the explicit values in Lemma~\ref{lem:overlaps}.
\end{proof}

The theorem makes precise both what the bare protocol does and does not guarantee: the diagonal of $\rho_{ID}$ is uniform, so the prescribed (computational-basis) use of the identity register leaks nothing; however, the off-diagonal terms do not vanish in general, so $\rho_{ID}$ is \emph{not} maximally mixed. Since the tallying procedure is repeated $O(KN^2\log(K/\delta))$ {times} (Section~\ref{subsec:measurement_complexity}), a Center deviating from the protocol could in principle estimate these coherences across repetitions and, combined with the publicly known tally, infer the equal-vote partition---which for $K=2$ with an unbalanced tally would identify each voter's choice. This motivates the following augmentation, which removes the off-diagonal terms altogether.

\medskip
\noindent\textbf{Protocol augmentation (identity dephasing).}
At the end of the voting stage and before tallying, the Center applies the diagonal unitary
\begin{equation}
    D_{\bm{\theta}} = \sum_{j} e^{i\theta_j} \ket{\id_j}\!\bra{\id_j},
    \label{eq:dephasing}
\end{equation}
where the phases $\theta_j$ are drawn independently and uniformly from $[0, 2\pi)$ and immediately discarded (never recorded). Operationally, $D_{\bm{\theta}}$ is realized by $n$ independent single-qubit phase rotations $e^{i\phi_q Z_q/2}$ on the identity qubits with $\phi_q$ drawn uniformly at random: since any two distinct identity states differ in at least one qubit, the induced phase differences $\theta_j - \theta_{j'}$ are uniformly distributed for all $j \neq j'$. Because every voting unitary $U_{\text{vote}}^{(j,k)}$ is block-diagonal in the identity basis (Eq.~\eqref{eq:vote_unitary}), $D_{\bm{\theta}}$ commutes with the entire voting stage and may equivalently be applied at any point before tallying. Moreover, $D_{\bm{\theta}}$ acts trivially on the ancilla and candidate registers, so \emph{all} prescribed measurement statistics---in particular the tally distribution of Eq.~\eqref{eq:vote_prob}---are unchanged.

\begin{theorem}[Maximal mixedness under identity dephasing]
\label{thm:dephasing}
With the identity-dephasing augmentation, the state of the full system averaged over the discarded phases is
\begin{equation}
    \bar{\rho} = \frac{1}{N} \sum_{j=0}^{N-1} \ket{\id_j}\!\bra{\id_j} \otimes \ket{\phi_j}\!\bra{\phi_j},
\end{equation}
and consequently the identity register is exactly maximally mixed on the $N$-dimensional voter subspace:
\begin{equation}
    \rho_{ID} = \frac{1}{N}\sum_{j=0}^{N-1} \ket{\id_j}\!\bra{\id_j},
    \qquad S(\rho_{ID}) = \log_2 N .
\end{equation}
\end{theorem}

\begin{proof}
Averaging $D_{\bm{\theta}}\ket{\Psi}\!\bra{\Psi}D_{\bm{\theta}}^\dagger$ over the phases, the off-diagonal blocks acquire the factor $\mathbb{E}\bigl[e^{i(\theta_j - \theta_{j'})}\bigr] = \delta_{jj'}$, which eliminates all $j \neq j'$ terms; the diagonal blocks are unaffected. Tracing out the ancilla and candidate registers then yields $\rho_{ID} = \mathbb{I}_N / N$ on the voter subspace.
\end{proof}

Finally, we state the operational anonymity guarantee, which holds for the (bare or augmented) protocol under semi-honest execution.

\begin{theorem}[Vote-permutation indistinguishability]
\label{thm:transcript}
Under semi-honest execution, the transcript of the protocol depends only on the tally vector $(n_0, \dots, n_{K-1})$. Specifically, in each repetition the ancilla outcome is a Bernoulli random variable with success probability $1/K$ independent of the vote assignment (Eq.~\eqref{eq:postselect_prob_repeat}), and, conditioned on a successful post-selection, the candidate-register outcome equals $\ket{c_k}\ket{c_k}$ with probability $n_k/N$ (Eq.~\eqref{eq:vote_prob}). Hence for any two vote assignments $v, v'$ with identical tallies, $T(v)$ and $T(v')$ are identically distributed, and the protocol satisfies Definition~\ref{def:anonymity}.
\end{theorem}

\begin{proof}
The repetitions are independent and identically distributed. Within each repetition, the ancilla outcome distribution is $(1 - 1/K,\, 1/K)$ by Eq.~\eqref{eq:postselect_prob_repeat}, which involves no dependence on the individual assignment $\{k_j\}$. Conditioned on the outcome $\ket{1}$, the post-selected state is Eq.~\eqref{eq:votes_diff_normalized}, and the candidate-register measurement yields $\ket{c_k}\ket{c_k}$ with probability $n_k/N$, a function of the tally alone. (If the ancilla reads $\ket{0}$, the prescribed protocol discards the round; any recorded outcome would similarly depend only on aggregate quantities.) The joint transcript distribution is therefore a function of $(n_0, \dots, n_{K-1})$ only.
\end{proof}

\paragraph{Adversary views and auxiliary information.}
Definition~\ref{def:anonymity} and Theorem~\ref{thm:transcript} concern the transcript of the prescribed measurements. Richer adversaries are conceivable, so we state precisely which \emph{views} are covered. Let $\mathrm{tally}(v)=(n_0,\dots,n_{K-1})$. We distinguish three views of one or more protocol runs:
\begin{itemize}
\item[\textbf{V1}] the classical transcript $T(v)$, together with arbitrary \emph{auxiliary information} $Z$ (the voter roll, prior beliefs, the ballots of colluding voters, \dots);
\item[\textbf{V2}] the quantum state of the identity register before the candidate measurement, for any number of runs;
\item[\textbf{V3}] any view that contains the identity register together with the candidate outcome, or the complete post-voting quantum state.
\end{itemize}

\begin{definition}[Anonymity with auxiliary information]
\label{def:anonymity_aux}
Let $Z$ be a random variable that may depend on $v$ in an arbitrary way and is independent of the protocol's internal randomness given $v$. The protocol is \emph{anonymous against $(T,Z)$} if $\Pr[T(v)=\tau\mid v,Z]=\Pr[T(v)=\tau\mid \mathrm{tally}(v)]$ for all $\tau$.
\end{definition}

\begin{corollary}[Auxiliary information]
\label{cor:aux}
Under the assumptions of Theorem~\ref{thm:transcript}, the protocol is anonymous against $(T,Z)$ for every $Z$. Consequently, for every prior on $v$, the posterior of $v$ given $(T(v),Z)$ equals the posterior given $(\mathrm{tally}(v),Z)$: the transcript can be simulated from the tally alone and leaks nothing beyond the published result.
\end{corollary}
\begin{proof}
By Theorem~\ref{thm:transcript} the law of $T(v)$ is a function of $\mathrm{tally}(v)$ only, hence $T(v)$ is conditionally independent of $(v,Z)$ given $\mathrm{tally}(v)$. The posterior statement follows from Bayes' rule.
\end{proof}
This is the strongest guarantee available for any protocol that publishes the tally: auxiliary information can still reveal votes through the tally itself, for example when all other ballots are known, or when the result is unanimous.

\begin{proposition}[Views beyond the transcript]
\label{prop:views}
Let the dephasing augmentation be used with fresh random phases in every run.
\begin{enumerate}
\item[(a)] \emph{Identity register alone (V2).} The joint state of the identity registers of any number $R$ of runs, before the candidate measurement, is $(\mathbb{I}_N/N)^{\otimes R}$, independent of $v$.
\item[(b)] \emph{Identity register with the outcome (V3).} Conditioned on ancilla outcome $\ket{1}$ and candidate outcome $k$, the identity register is in the state $\frac{1}{n_k}\sum_{j:\,k_j=k}\ket{\id_j}\!\bra{\id_j}$; measuring it returns a uniformly random voter among the $n_k$ voters who voted for $k$.
\item[(c)] \emph{Complete state (V3).} Measuring the identity register of the post-dephasing state $\bar\rho$ returns a uniformly random voter $j$ and leaves the ancilla--candidate registers in $\ket{\phi_{k_j}}$. The square-root measurement on $\{\ket{\phi_k}\}_k$ then identifies $k_j$ with probability $P_K=\frac{1}{K^2}\bigl(\sqrt{\lambda_1}+(K-1)\sqrt{\lambda_2}\bigr)^2$, where $\lambda_1=1+\frac{(K-1)(K-2)}{K}$ and $\lambda_2=\frac{2}{K}$. In particular $P_2=1$, $P_3=0.95$, and $P_4\approx0.86$.
\end{enumerate}
\end{proposition}
\begin{proof}
(a) Each run uses an independent replay and independent phases, so the registers are independent, and each equals $\mathbb{I}_N/N$ by Theorem~\ref{thm:dephasing}. (b) After the ancilla outcome $\ket{1}$ the state is proportional to $\sum_j\ket{\id_j}\ket{c_{k_j}}_A\ket{c_{k_j}}_B$ (Eq.~\eqref{eq:votes_diff_normalized}); the candidate outcome $k$ projects onto the terms with $k_j=k$, and the dephasing removes the coherences between them. (c) By Lemma~\ref{lem:overlaps} the Gram matrix of $\{\ket{\phi_k}\}_k$ has diagonal $1$ and off-diagonal entries $(K-2)/K$, with eigenvalues $\lambda_1$ (once) and $\lambda_2$ ($K-1$ times). For equiprobable states with this symmetry the square-root measurement succeeds with probability $\frac{1}{K^2}(\sqrt{\lambda_1}+(K-1)\sqrt{\lambda_2})^2$; for $K=2$ the states are orthogonal.
\end{proof}

\paragraph{Scope of the anonymity claim.}
Anonymity is therefore claimed against V1 with arbitrary auxiliary information (Corollary~\ref{cor:aux}) and against V2 (Proposition~\ref{prop:views}(a)). It is \emph{not} claimed against V3: the complete quantum state of the machine, or the identity register together with the candidate outcome, reveals the vote of a random voter in every run (Proposition~\ref{prop:views}(b),(c)), and replays reveal further voters. This is not an accident of the analysis, since the classical correlation between identity and vote registers is what enforces one vote per voter. View V3 coincides with the trusted base of assumption M4: the machine's internal state, including its stored records, must not be exposed, exactly as for a classical voting machine. In the distributed setting the Center holds only part of the global state. Because that global state is not derived (D4), we make no claim for the Center's complete quantum view; the semi-honest Center is assumed to obtain V1 only.

\subsection{Resilience Against Specific Attacks}
\label{subsec:attacks}

\paragraph{Identity Forgery (External Eavesdropper)}
In the distributed setting, an eavesdropper Eve could intercept the qubit sent to the machine serving voter $V_j$ and substitute her own. This is prevented by the authenticated quantum channel assumption. Even if the channel is not authenticated, any measurement or substitution by Eve disturbs the entanglement between the Center and the machine. The disturbance is detected by the CHSH-based verification protocol described in Section~\ref{subsec:verification} {(Theorem~\ref{thm:soundness}, for the tested pairs and without an i.i.d.\ assumption)}. The no-cloning theorem guarantees that Eve cannot perfectly copy the qubit without introducing detectable errors.

\paragraph{One Vote per Voter}
Each voter's identity branch is processed exactly once, by the machine, under the Center's coordination, and voters {cannot apply operations beyond the ballot interface (M2)}. Double voting is therefore prevented by the machine's ballot handling, as in a classical voting machine; this is a trust assumption (M1--M2 in Section~\ref{sec:models}) and not a consequence of entanglement.

\paragraph{Mistaken or Faulty Ballots}
A voter may select a wrong candidate (a mistaken vote), and a machine may execute a ballot operation incorrectly (for example, through a miscalibrated gate). A mistaken selection is a valid vote and is indistinguishable from an intended one. A faulty operation $\tilde{U} \neq Z_k$ produces the difference state
\begin{equation}
    \ket{\tilde{\psi}} - \ket{\psi_\cands},
\end{equation}
which is generally not confined to a single candidate's basis state. Part of its weight may fall on outcomes outside the valid candidate states (``spoiled'' outcomes), but a faulty operation that stays within the valid subspace, for example an additional phase, produces no spoiled outcome. We therefore make no claim about the influence of a single faulty ballot: after post-selection, a faulty operation can carry more weight than a correct vote, which has weight $1/K$ in the post-selected pool. The numerical effect is shown in Section~\ref{subsec:dishonest}.

\subsection{Entanglement Verification}
\label{subsec:verification}

The entanglement shared between the machines and the Center can be verified: an unauthorized measurement or substitution of the distributed qubits produces a statistically detectable drop in a Bell-test statistic. The check concerns the entanglement and not the ballots. We state the threat model and the protocol, and then establish completeness (honest noisy hardware passes) and soundness (a tampered distribution is rejected) separately in Section~\ref{subsec:finite_stats}, without an i.i.d.\ assumption on the attack.

\paragraph{Threat model.} The adversary (A1) acts on the quantum channels during distribution with arbitrary operations on all transmitted qubits: coherent across pairs, adaptive, not i.i.d., and with an arbitrary quantum memory. The measurement devices of the Center and the machines are trusted to implement the qubit measurements stated below (a device-dependent test), the classical channels are authenticated, and the machines store their qubits reliably after delivery (D2).

\paragraph{Protocol.}
\begin{enumerate}
\item \emph{Distribution.} The Center prepares $M=N+n_t$ entangled pairs and sends the machine-side qubits through the channels. Test pairs are Bell pairs $\ket{\Phi^+}=(\ket{00}+\ket{11})/\sqrt2$. Each machine acknowledges receipt over the authenticated classical channel.
\item \emph{Unpredictable selection after distribution.} Only after all acknowledgements have arrived, the Center draws from fresh private randomness a uniformly random subset $S$ of $n_t$ pairs as test pairs (the other $N$ pairs are used for voting), and for each test pair $t\in S$ a uniformly random setting $x_t\in\{0,1\}$. The machine draws $y_t\in\{0,1\}$ uniformly at random. The subset and the settings are therefore independent of everything the adversary has done.
\item \emph{Measurement.} For each $t\in S$ the Center measures $A_{x_t}$ and the machine measures $B_{y_t}$, where
\begin{equation*}
A_0=Z,\quad A_1=X,\quad B_0=\tfrac{Z+X}{\sqrt2},\quad B_1=\tfrac{Z-X}{\sqrt2},
\end{equation*}
with outcomes $a_t,b_t\in\{0,1\}$. After both measurements are completed, the settings and outcomes are announced over the authenticated channel.
\item \emph{Statistic and decision.} For each test pair compute $Y_t=4\,\sigma(x_t,y_t)\,(-1)^{a_t\oplus b_t}$ with $\sigma(1,1)=-1$ and $\sigma(x,y)=+1$ otherwise, and $\widehat S=\frac1{n_t}\sum_{t\in S}Y_t$. The protocol aborts if $\widehat S<\tau$, for a threshold $2<\tau<2\sqrt2$.
\end{enumerate}
Because the settings are uniform and independent of the state of the pair, $\mathbb{E}[Y_t\mid\text{state of pair }t]=\langle B\rangle$, where $B=A_0\otimes(B_0+B_1)+A_1\otimes(B_0-B_1)$ is the Bell operator, that is, the CHSH combination $\langle A_0B_0\rangle+\langle A_0B_1\rangle+\langle A_1B_0\rangle-\langle A_1B_1\rangle$. Each $Y_t$ lies in $[-4,4]$.

\begin{lemma}[Bell-operator witness]
\label{lem:witness}
For the settings above, $B=2\sqrt2\bigl(\ket{\Phi^+}\!\bra{\Phi^+}-\ket{\Psi^-}\!\bra{\Psi^-}\bigr)$. Hence, for every two-qubit state $\rho$, $\operatorname{Tr}(B\rho)\le 2\sqrt2\,\langle\Phi^+|\rho|\Phi^+\rangle\le 2\sqrt2$, and every separable state satisfies $\operatorname{Tr}(B\rho)\le2$.
\end{lemma}
\begin{proof}
With these settings $B=\sqrt2\,(Z\otimes Z+X\otimes X)$. The Bell states $\Phi^\pm,\Psi^\pm$ are eigenvectors of $Z\otimes Z+X\otimes X$ with eigenvalues $2,0,0,-2$ for $\Phi^+,\Phi^-,\Psi^+,\Psi^-$. The separable bound is the CHSH (local-hidden-variable) bound~\cite{clauser1969chsh}.
\end{proof}

\subsection{Completeness, Soundness, and Finite-Size Analysis}
\label{subsec:finite_stats}

Let $\mathcal F_{t-1}$ denote everything that has happened before test round $t$ (the adversary's actions, all earlier settings and outcomes), let $\rho_t$ be the state of the $t$-th tested pair conditioned on $\mathcal F_{t-1}$, and let $s_t=\operatorname{Tr}(B\rho_t)=\mathbb{E}[Y_t\mid\mathcal F_{t-1}]$ and $F_t=\langle\Phi^+|\rho_t|\Phi^+\rangle$. The differences $Y_t-s_t$ form a martingale difference sequence with values in an interval of length $8$, with no assumption that the rounds are independent or identically distributed. We use the Azuma--Hoeffding inequality~\cite{azuma1967weighted}: for such a sequence, $\Pr\bigl[\sum_{t}(Y_t-s_t)\ge n_t\varepsilon\bigr]\le e^{-n_t\varepsilon^2/32}$, and the same bound holds for the lower tail.

\begin{theorem}[Completeness]
\label{thm:completeness}
Suppose honest noisy hardware satisfies $s_t\ge s_h$ in every round (for example, Werner pairs of visibility $v$ give $s_t=2\sqrt2\,v$), and let $\tau<s_h$. Then the protocol aborts with probability at most $\exp\bigl(-n_t(s_h-\tau)^2/32\bigr)$.
\end{theorem}
\begin{proof}
Abort means $\widehat S<\tau$, which implies $\frac1{n_t}\sum_t(Y_t-s_t)<\tau-s_h$ since $\frac1{n_t}\sum_ts_t\ge s_h$. Apply the lower-tail bound.
\end{proof}

\begin{theorem}[Soundness for the tested pairs]
\label{thm:soundness}
Let $\bar F=\frac1{n_t}\sum_tF_t$ be the average conditional fidelity of the tested pairs with $\ket{\Phi^+}$. For every attack (adaptive, coherent across pairs, not i.i.d.) and every $\varepsilon>0$,
\begin{equation}
\Pr\Bigl[\widehat S\ge\tau\ \text{and}\ \bar F<\tfrac{\tau-\varepsilon}{2\sqrt2}\Bigr]\le e^{-n_t\varepsilon^2/32}.
\end{equation}
\end{theorem}
\begin{proof}
By Lemma~\ref{lem:witness}, $s_t\le2\sqrt2\,F_t$. If $\bar F<(\tau-\varepsilon)/(2\sqrt2)$ then $\frac1{n_t}\sum_ts_t<\tau-\varepsilon$, so $\widehat S\ge\tau$ implies $\frac1{n_t}\sum_t(Y_t-s_t)>\varepsilon$. Apply the upper-tail bound.
\end{proof}
By Uhlmann's theorem and the relation between fidelity and trace distance~\cite{nielsen2010quantum}, a pair with $F_t\ge1-\eta$ has joint state with any system $E$ held by the adversary within trace distance $\sqrt\eta$ of $\ket{\Phi^+}\!\bra{\Phi^+}\otimes\sigma_E$. Thus a passed test limits what the adversary can know about the tested pairs, even if its attack was not i.i.d.

\paragraph{Why the selection must be unpredictable.} The martingale argument uses that $(x_t,y_t)$ is uniform and independent of $\rho_t$. If the adversary learns the subset $S$ in advance, it leaves the test pairs untouched and attacks only the voting pairs, and the test is vacuous. If it learns $(x_t,y_t)$ before acting on pair $t$, it can deliver a product state whose outcomes satisfy the CHSH condition for exactly that setting, which gives $Y_t=4$ in every round and passes the test with certainty. This is why $S$ and the settings are drawn only after all acknowledgements, from fresh randomness (for example a quantum random number generator or a public randomness beacon), and why the announcement follows the measurements.

\paragraph{Finite-size example.} The honest-case and tamper-case errors are both below $\epsilon$ if $n_t\ge\frac{32}{\min\{s_h-\tau,\ \varepsilon_s\}^2}\ln\frac1{\epsilon}$, with $\varepsilon_s$ the slack in Theorem~\ref{thm:soundness}. For Werner pairs of visibility $v=0.95$ ($s_h=2.687$) and $\tau=2.49$, $\varepsilon_s=0.2$, $\epsilon=10^{-3}$, this gives $n_t\approx5.7\times10^{3}$ test pairs, and a passed test certifies an average conditional fidelity $\bar F\ge0.81$ of the tested pairs; with $\tau=2.4$, $\varepsilon_s=0.3$ it gives $n_t\approx2.7\times10^3$ and $\bar F\ge0.74$. Every tallying run consumes a fresh set of pairs (D3), so the test is performed per batch; to keep the overall error below $\epsilon$ over $B$ batches, replace $\ln(1/\epsilon)$ by $\ln(B/\epsilon)$.

\paragraph{From tested pairs to voting pairs.} Theorem~\ref{thm:soundness} concerns the pairs that were measured. For the voting pairs, two statements are relevant. (i) Since $S$ is a uniformly random subset, for any joint state of the $M$ pairs, $\mathbb{E}_S[\widehat S]=\frac1M\sum_{i}\langle B_i\rangle=\mathbb{E}_S\bigl[\frac1N\sum_{i\notin S}\langle B_i\rangle\bigr]$, so the test statistic is an unbiased estimator of the average Bell value of the untested pairs. (ii) A high-probability statement, namely that passing the test implies that the voting pairs are close to ideal pairs except with small probability, requires a sampling argument for quantum populations (see, for example, Ref.~\cite{bouman2010sampling}) or a de Finetti-type argument. We do not prove it here, and we state it as an explicit assumption: the passed test is taken to certify the voting pairs. The same assumption connects the Bell test pairs to the candidate pairs used for voting when the candidate state differs from $\ket{\Phi^+}$ ($K>2$).

\paragraph{Scope.} The test detects tampering with the entanglement during distribution, under the threat model above. It does not detect misbehavior of the machines or the Center, and it does not establish the integrity of the election by itself.

\section{Numerical Experiments}
\label{sec:numerical}

To validate the theoretical claims and assess the practical behavior of the proposed quantum voting protocol, we conducted numerical simulations using PennyLane \cite{bergholm2018pennylane}. The circuits were executed both with exact statevector simulation (ideal case) and with shot‑based sampling to model finite‑statistics effects. We considered the two illustrative scenarios from Section~\ref{sec:examples}: 4 voters with 2 candidates (Blue vs.\ Red) and 8 voters with 3 candidates. For each scenario we examined three cases: ideal (no noise, honest voters), depolarizing noise on the candidate qubits, and {faulty ballots in which the prescribed phase flip is replaced by an arbitrary unitary operation}. Finally, we studied the convergence of the tallying probabilities as a function of the number of protocol repetitions (shots). {Section~\ref{subsec:extended} then extends the study to larger instances, memory, readout, and gate noise, and confidence intervals.}

\subsection{Ideal Case}
\label{subsec:ideal}

\begin{figure}[htbp]
    \centering
    \includegraphics[width=\columnwidth]{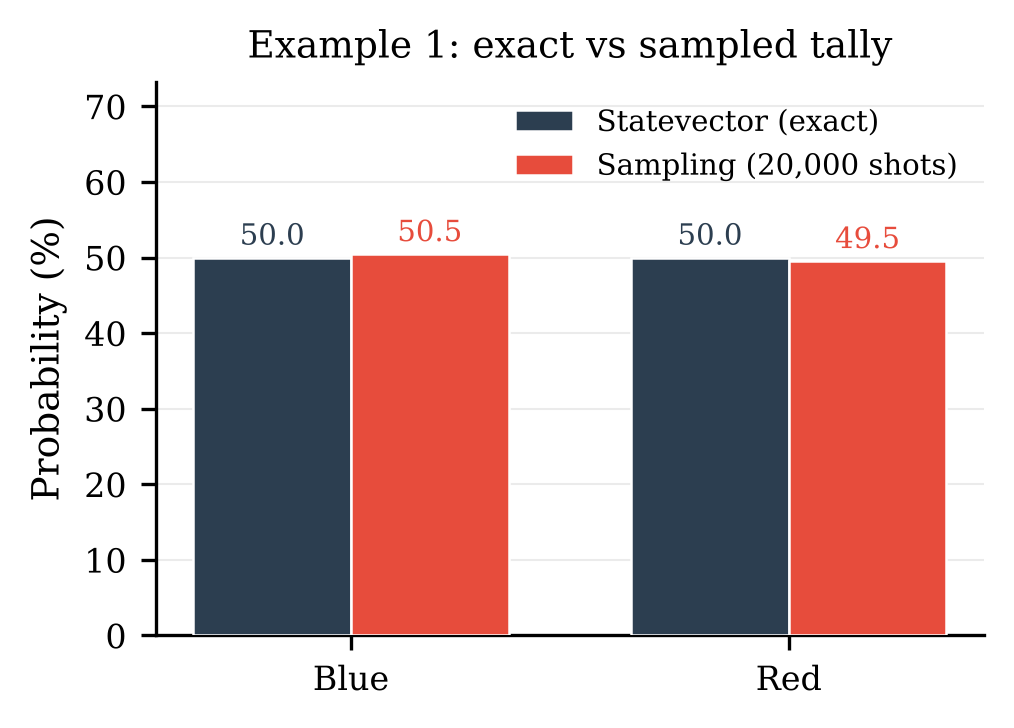}
    \caption{Example 1 (4 voters, 2 candidates): ideal case. Statevector (exact) vs.\ sampling ($20\,000$ shots). The Blue and Red probabilities are both $50\%$ as expected.}
    \label{fig:ex1_ideal}
\end{figure}

\begin{figure}[htbp]
    \centering
    \includegraphics[width=\columnwidth]{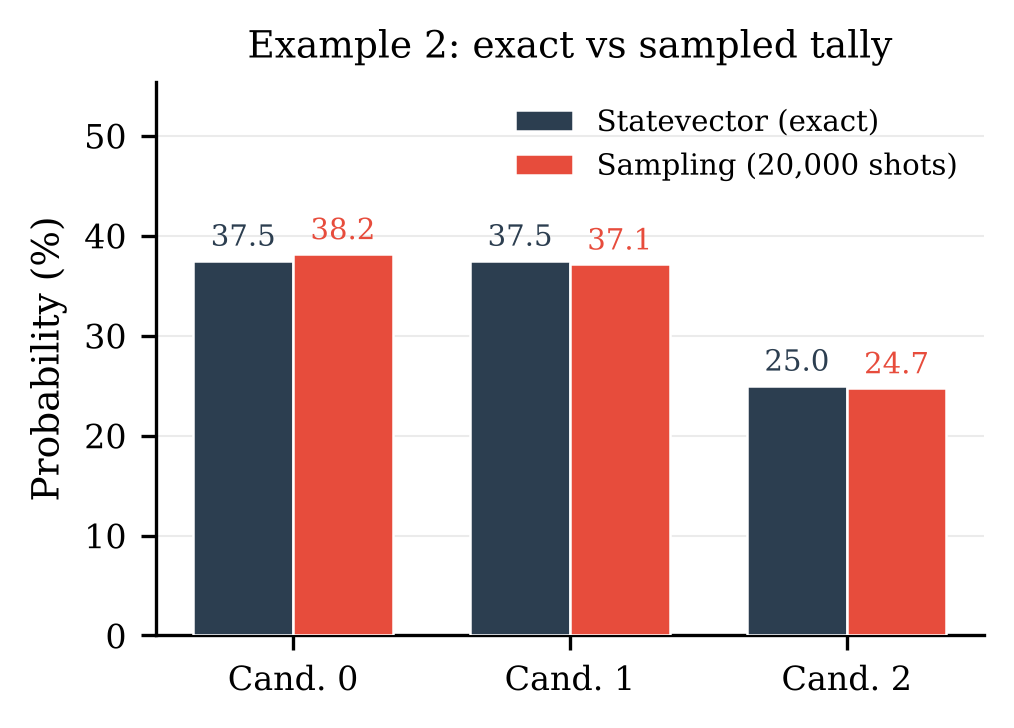}
    \caption{Example 2 (8 voters, 3 candidates): ideal case. The theoretical probabilities $37.5\%$, $37.5\%$, $25.0\%$ are faithfully reproduced by the sampling.}
    \label{fig:ex2_ideal}
\end{figure}

Under ideal conditions (no noise, all voters honest), the protocol exactly reproduces the theoretical vote distributions derived in Section~\ref{sec:examples}. Figures~\ref{fig:ex1_ideal} and~\ref{fig:ex2_ideal} compare the statevector (exact) probabilities with the empirical probabilities obtained from $20\,000$ shots. In both cases the sampling results match the theoretical values within statistical fluctuations, confirming the correctness of the voting and tallying mechanisms.

\subsection{Effect of Depolarizing Noise}
\label{subsec:noise}

\begin{figure}[htbp]
    \centering
    \includegraphics[width=\columnwidth]{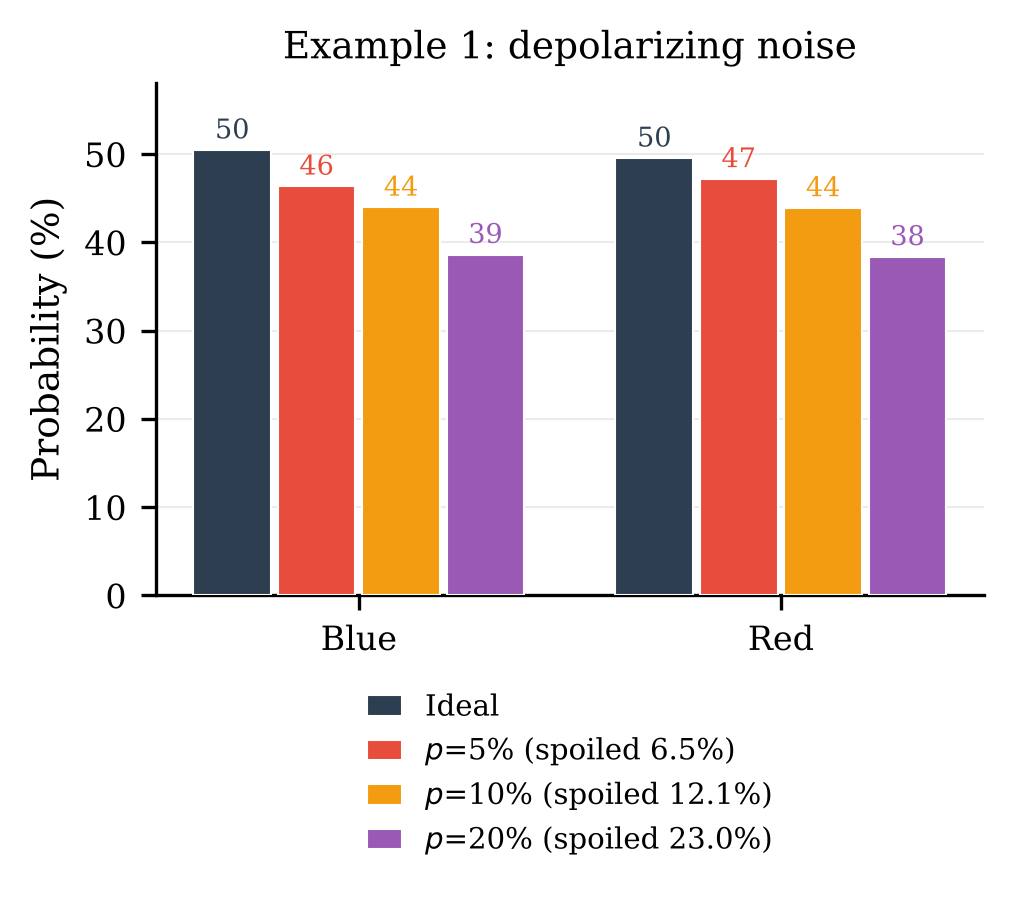}
    \caption{Example 1 with depolarizing noise. Spoiled votes appear as the noise level increases, and the Blue/Red probabilities deviate from $50\%$.}
    \label{fig:ex1_noise}
\end{figure}

\begin{figure}[htbp]
    \centering
    \includegraphics[width=\columnwidth]{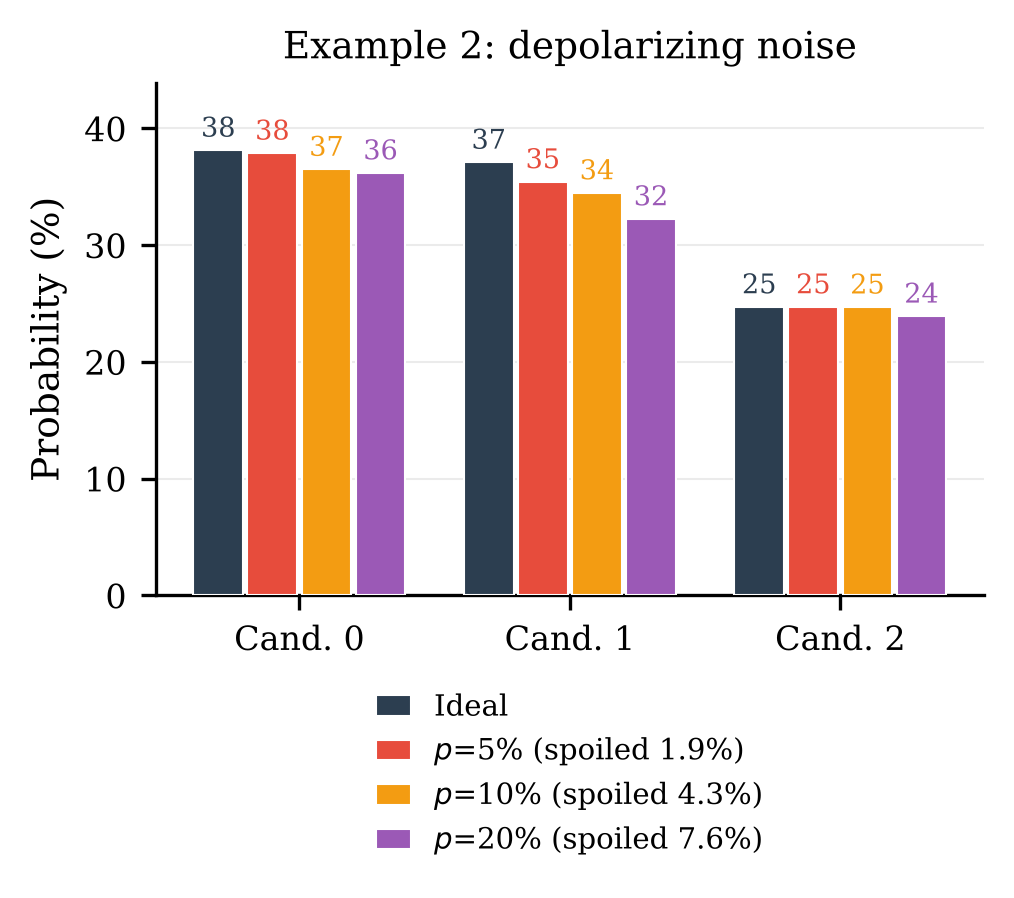}
    \caption{Example 2 with depolarizing noise. The unused basis state $\ket{11}$ becomes a spoiled indicator; higher noise leads to more spoiled votes and a distortion of the tally.}
    \label{fig:ex2_noise}
\end{figure}

In realistic quantum hardware, decoherence introduces errors. We modeled this by applying a depolarizing channel with probability $p$ to each candidate qubit after the voting stage. The noise causes probability mass to leak into basis states that do not correspond to any valid candidate (``spoiled votes''). Figures~\ref{fig:ex1_noise} and~\ref{fig:ex2_noise} show the results for $p = 5\%, 10\%, 20\%$. As $p$ increases, the fraction of spoiled votes grows, and the distribution among the legitimate candidates becomes distorted. This behavior is consistent with the expected degradation of the quantum state under noise. {No error correction or mitigation is applied, so the noise distorts the tally; we do not claim robustness against such errors beyond the margin analysis of Section~\ref{subsec:extended}.}

\subsection{Faulty or Mistaken Ballots}
\label{subsec:dishonest}

\begin{figure}[htbp]
    \centering
    \includegraphics[width=\columnwidth]{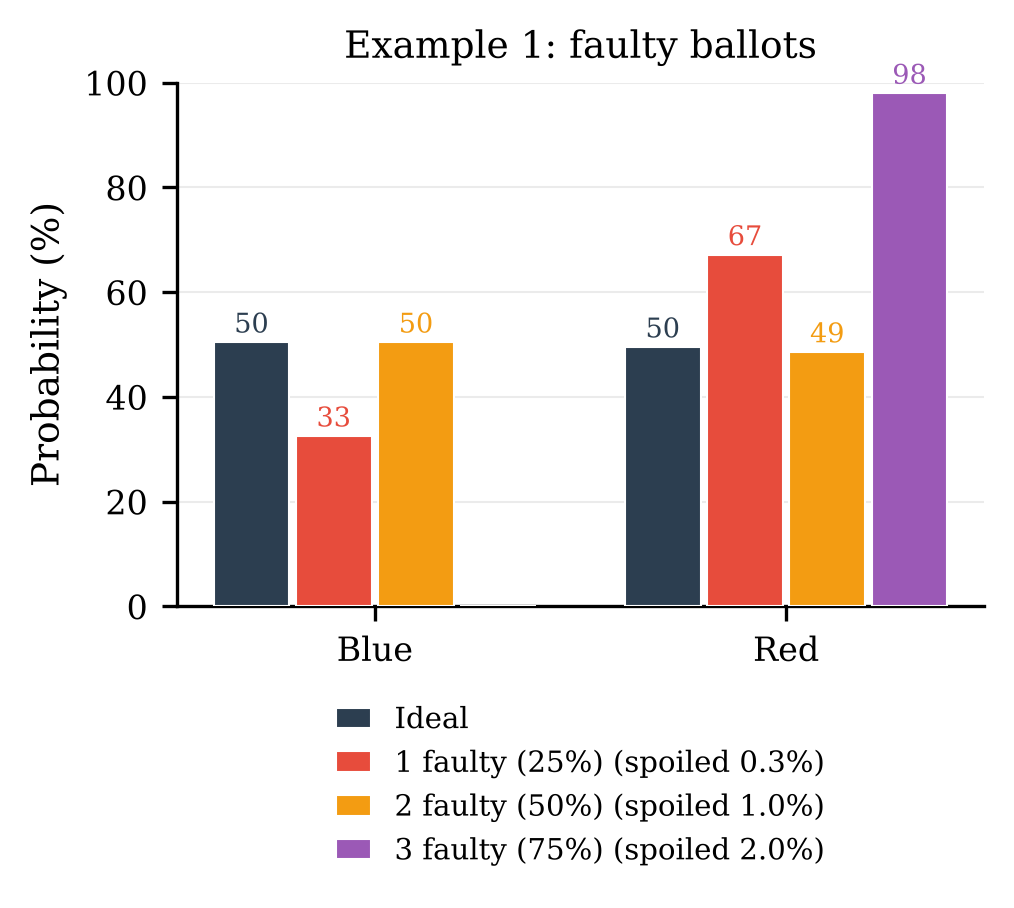}
    \caption{Example 1 with {faulty ballots. As the number of faulty ballots grows}, the vote distribution deviates from the ideal $50\%$–$50\%$ split and spoiled votes appear.}
    \label{fig:ex1_dishonest}
\end{figure}

\begin{figure}[htbp]
    \centering
    \includegraphics[width=\columnwidth]{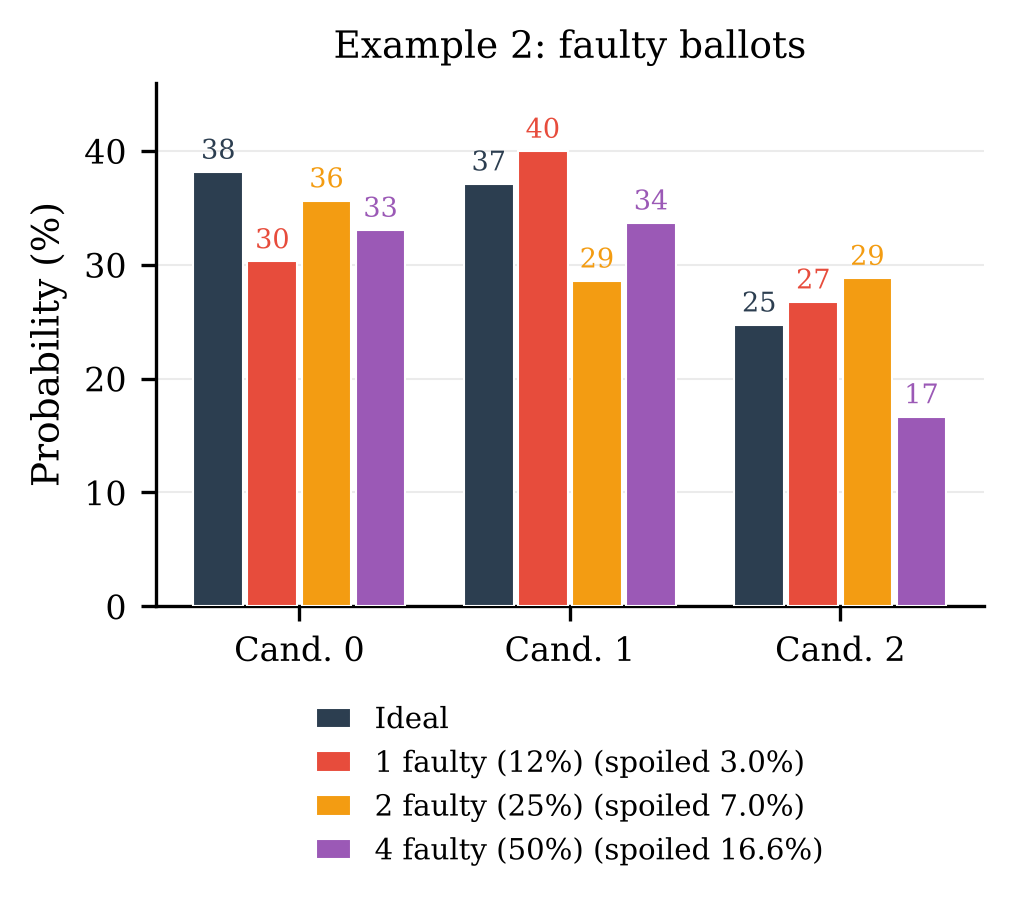}
    \caption{Example 2 with {faulty ballots. Even a single faulty ballot ($12.5\%$ of the ballots)} already causes noticeable distortion and spoiled votes.}
    \label{fig:ex2_dishonest}
\end{figure}

To test the effect of faulty ballot operations, we replaced the honest phase-flip operation $Z_k$ by an arbitrary unitary $U = R_Y(\pi/3) \otimes R_Y(\pi/4)$ on the two candidate qubits; this models, for example, a miscalibrated gate in the machine. Figures~\ref{fig:ex1_dishonest} and~\ref{fig:ex2_dishonest} show the effect of increasing the number of faulty ballots. The tally becomes significantly distorted, and spoiled votes appear. We make no claim about the influence of a single faulty ballot: in Example 1, for instance, one faulty ballot out of four (25\%) moves the tally from $50\%/50\%$ to about $33\%/67\%$. Spoiled outcomes flag only those faults that leave the valid candidate subspace; faults that act within it, for example an additional phase on a candidate state, produce no spoiled outcome.

\subsection{Sampling Convergence}
\label{subsec:convergence}

\begin{figure}[htbp]
    \centering
    \includegraphics[width=\columnwidth]{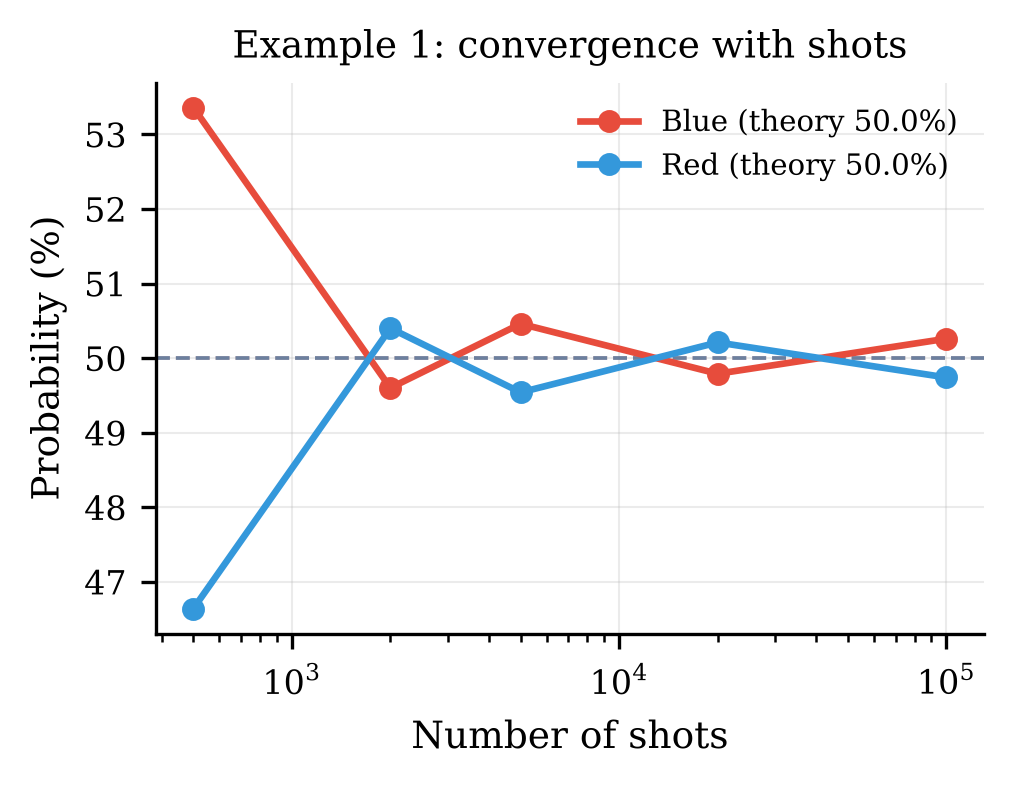}
    \caption{Convergence of tallying probabilities for Example 1 (tie between Blue and Red). The empirical difference between the two candidates decreases as the number of shots increases, following the expected statistical behavior.}
    \label{fig:ex1_convergence}
\end{figure}

\begin{figure}[htbp]
    \centering
    \includegraphics[width=\columnwidth]{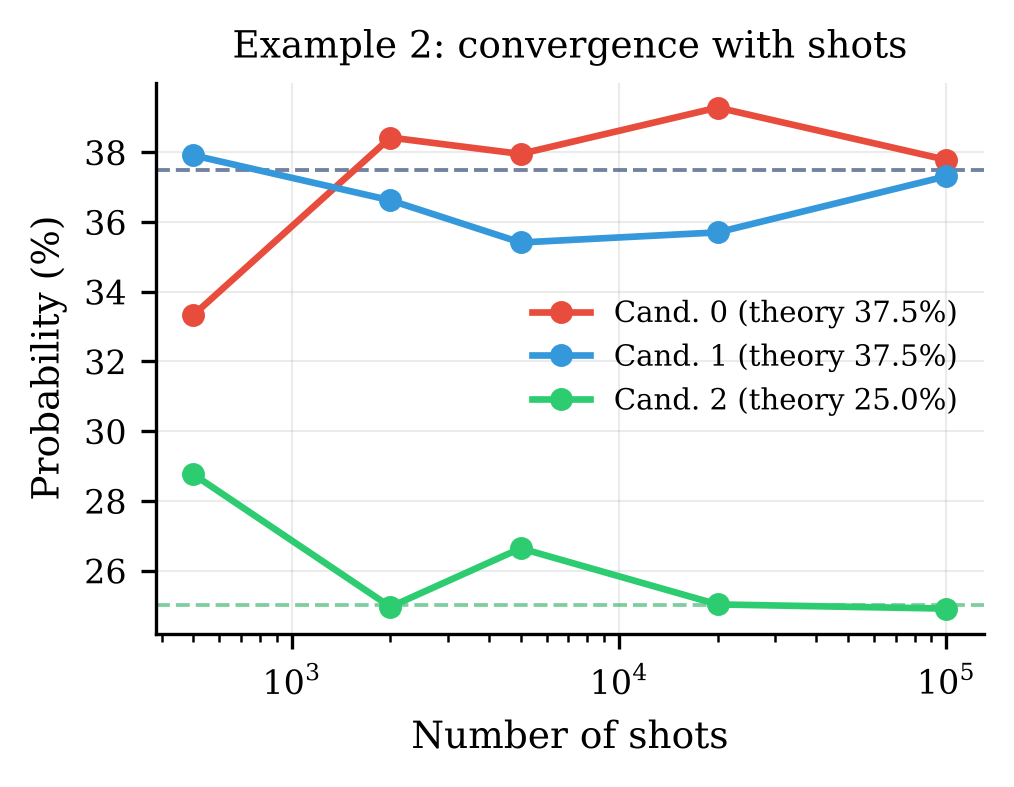}
    \caption{Convergence of tallying probabilities for Example 2. The three candidates’ probabilities approach the theoretical values $37.5\%$, $37.5\%$, $25.0\%$ as the number of shots grows.}
    \label{fig:ex2_convergence}
\end{figure}

The tallying procedure relies on post‑selection on the ancilla qubit, which succeeds with probability $1/K$ (Eq.~\eqref{eq:postselect_prob_repeat}). Consequently, multiple protocol executions (shots) are required to accumulate enough successful measurements. We studied the convergence of the empirical probabilities as a function of the number of shots. Figures~\ref{fig:ex1_convergence} and~\ref{fig:ex2_convergence} plot the measured probabilities against shot count on a logarithmic scale, together with the theoretical exact values. The empirical probabilities converge rapidly, and the fluctuations are {consistent with the concentration bounds (Hoeffding and Bernstein) derived in Section~\ref{subsec:measurement_complexity}}. In the 4‑voter, 2‑candidate case (a perfect tie), the difference between Blue and Red shrinks as $1/\sqrt{R}$, consistent with the statistical uncertainty of a fair coin. In the 8‑voter, 3‑candidate case, the probabilities converge to their respective theoretical fractions.

\subsection{{Extended Simulations: Larger Instances, Noise Models, and Uncertainty}}
\label{subsec:extended}

The examples above are small and use a single noise model. To address larger instances, additional noise sources, and statistical uncertainty, we repeated the study with an independent NumPy simulator (code and results are provided with the paper; see the data availability statement). It reproduces the PennyLane results of the examples above (for Example 1 the spoiled fractions at $p=5\%,10\%,20\%$ are $6.4\%,12.4\%,23.1\%$, against $6.5\%,12.1\%,23.0\%$ in the sampled results above) and the closed-form tally probabilities up to machine precision for up to $N=256$ voters and $K=8$ candidates. Noise that acts on the control and candidate registers after the voting stage can be simulated exactly for any $N$: tracing out the identity register leaves the mixture $\sum_k \frac{n_k}{N}\,\mathcal{E}(\ket{\chi_k}\!\bra{\chi_k})$ of the single-voter states, which allows $N$ up to $4096$. Errors that occur \emph{during} voting can move amplitude between identity branches; these are simulated with full state vectors up to $N=128$ (seven identity qubits) by Pauli-error trajectories.

\paragraph{Noise models and metric.} (i) Depolarizing noise on each candidate qubit after voting, as in Section~\ref{subsec:noise}. (ii) Memory noise: dephasing or amplitude damping with probability $\epsilon$ per voting step on the control and candidate qubits, composed over the $N$ steps, since the candidate register idles during all $N$ ballot operations. (iii) Readout error: a bit flip with probability $r$ on every measured bit. (iv) Gate errors: each ballot operation is a multi-controlled gate decomposed into $g(n)=12n+6$ two-qubit gates; each gate fails with probability $p_2$, and a failure applies a uniformly random Pauli error to a uniformly random identity, control, or candidate qubit (a Pauli-twirled model; the auxiliary qubits of the decomposition are not modeled). Each point uses $300$ trajectories, and intervals are $95\%$ bootstrap intervals. The metric is the \emph{retained margin} $\rho=(q_1-\max_{j\ne1}q_j)/(p_1-\max_{j\ne1}p_j)$, where $q$ are the shares among the accepted outcomes under noise and $p$ the true shares. Then $\rho=1$ without noise and $\rho\le0$ means that the decision flips. Spoiled outcomes count for no candidate, so Proposition~\ref{prop:winner} applies with $\Delta$ replaced by $\rho\Delta$, and the number of accepted outcomes needed for winner identification grows by the factor $\rho^{-2}$.

\begin{figure*}[htbp]
    \centering
    \includegraphics[width=0.95\textwidth]{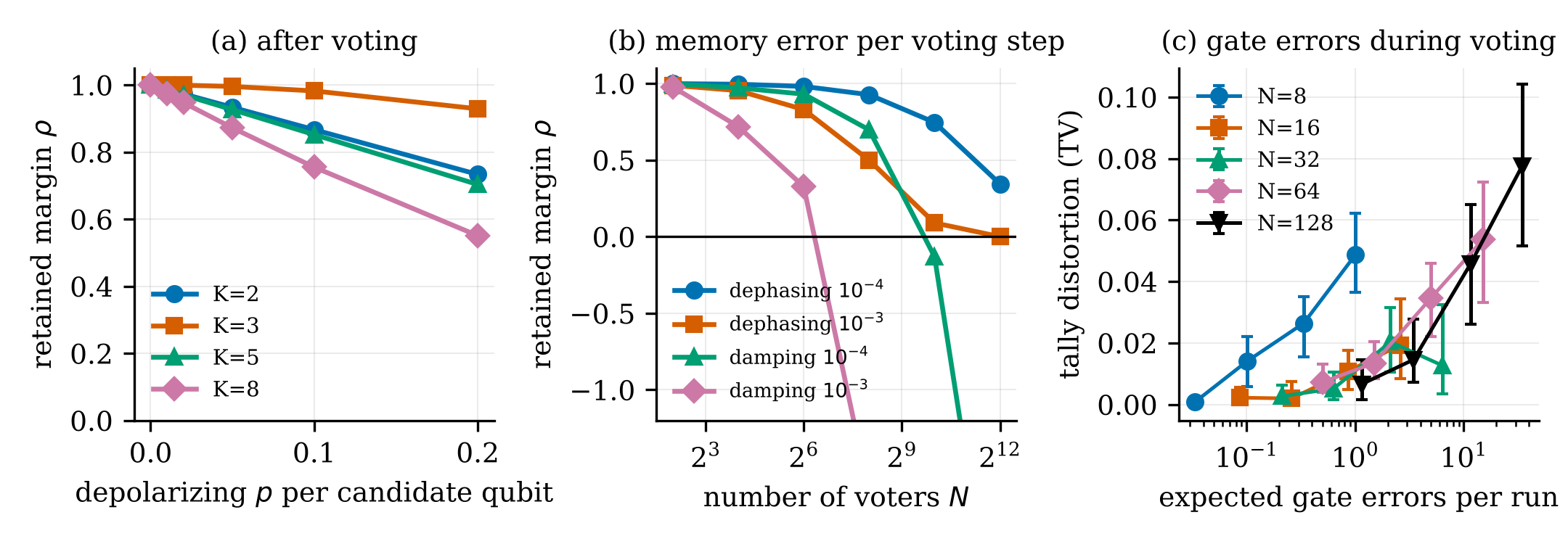}
    \caption{{Effect of noise on a race with margin $\Delta=0.1$. (a) Depolarizing noise after voting, $N=100$, for $K=2,3,5,8$. (b) Memory noise per voting step, $K=3$, with the leader on the candidate that is \emph{not} labeled $\ket{0\dots0}$; $\rho<0$ means a wrong winner. (c) Total-variation distortion of the tally under gate errors during voting ($K=3$, Pauli trajectories, $95\%$ bootstrap intervals), against the expected number of gate errors per run $g(n)\,p_2\,N$.}}
    \label{fig:noise_margin}
\end{figure*}

\begin{figure*}[htbp]
    \centering
    \includegraphics[width=0.95\textwidth]{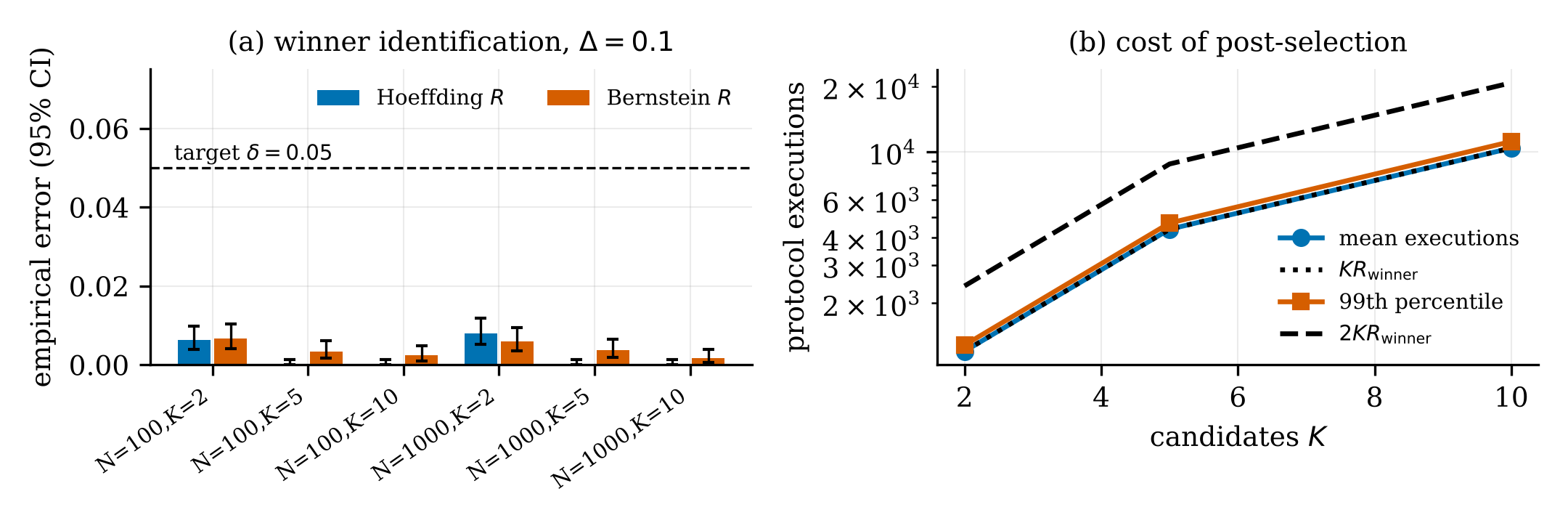}
    \caption{{Check of the sampling bounds ($3000$ repetitions each, $\delta=0.05$, $\Delta=0.1$). (a) Empirical winner-identification error with $95\%$ Clopper--Pearson intervals when $R$ equals the Hoeffding size of Eq.~\eqref{eq:R_winner} or the Bernstein size of Eq.~\eqref{eq:R_winner_bern}. (b) Number of protocol executions needed to collect $R_{\text{winner}}$ accepted outcomes ($N=1000$).}}
    \label{fig:sampling_check}
\end{figure*}

\paragraph{Sampling bounds (Figure~\ref{fig:sampling_check}).} For $N\in\{100,1000\}$ and $K\in\{2,5,10\}$ the winner-identification error with the Hoeffding size was between $0.6\%$ and $0.8\%$ for $K=2$ (upper confidence limit $1.2\%$) and $0$ for $K=5,10$ (upper limit $0.12\%$); the Bernstein sizes, which are smaller for $K=5,10$ ($436$ and $329$ instead of $877$ and $1039$), gave $0.2\%$ to $0.7\%$ (upper limit $1.0\%$). All are below $\delta=0.05$, so the bounds are valid and conservative. The expected number of executions to collect $R$ accepted outcomes equals $KR$, and its $99$th percentile stayed below $2KR$. The anytime rule of Section~\ref{subsubsec:adaptive} made no wrong decision in $600$ runs per setting (upper limit $0.6\%$) and used $3.7$ to $4.0$ times the fixed-sample size. For exact tallying with $N=10$ to $20$ and $K=3,5$, the error with $R=2N^2\ln(2K/\delta)$ was between $0.03\%$ and $0.3\%$.

\paragraph{Noise (Figure~\ref{fig:noise_margin}).} Depolarizing noise after voting does not depend on $N$. It kept the winner correct for all $p\le0.2$ and $K=2,\dots,8$ and reduced the retained margin to $\rho=0.73,0.93,0.70,0.55$ for $K=2,3,5,8$ at $p=0.2$ (with spoiled fractions $23\%$, $7.6\%$, $11\%$, and $0$, the last because all eight basis states of three qubits are candidates). Readout error up to $r=5\%$ per bit reduced $\rho$ to $0.87$ for $K=3$ and added $3\%$ spoiled outcomes. Memory noise is different, because it accumulates during the $N$ voting steps: with $\epsilon=10^{-4}$ per step, dephasing keeps $\rho\ge0.92$ up to $N=256$ and gives $\rho=0.75$ at $N=1024$, whereas with $\epsilon=10^{-3}$ it gives $\rho=0.50$ at $N=256$, $0.09$ at $N=1024$, and no margin at $N=4096$. Amplitude damping is worse and, in addition, biased: it drives the register towards $\ket{0\dots0}$, which favors the candidate with that label; with the leader on a different label the decision flips at $N=256$ for $\epsilon=10^{-3}$ and at $N=1024$ for $\epsilon=10^{-4}$. The required memory coherence therefore grows linearly with $N$ times the duration of one ballot operation. Gate errors during voting distort the tally through spoiled outcomes and through errors on identity qubits, which move later ballots to wrong branches. For $N=128$ the distortion is below $1.5\%$ (upper interval limit) at about one error per run, the winner remained correct up to about $11$ errors per run ($\rho=0.24$), and the decision flipped at about $35$ errors per run. A tally close to the ideal one thus requires $g(n)\,p_2\,N\lesssim1$, i.e.\ $p_2\lesssim10^{-4}$ at $N=128$.

\paragraph{Limitations.} All results are simulations. No hardware experiment was performed. The circuit size is $N\,g(n)$ two-qubit gates per run ($\approx10^2$ for $N=4$ and $\approx10^4$ for $N=128$), and by the above the tally at larger $N$ requires gate and memory error rates well below those of current devices or error correction. The amplitude-damping bias suggests randomizing the assignment of candidates to basis states between replays, which we have not tested. The simulations do not model errors of the Hadamard test beyond the readout model, correlated noise, or leakage.

{The numerical experiments thus confirm the theoretical analysis in the ideal case: the protocol reproduces the tally probabilities and exhibits statistical convergence consistent with the sampling analysis. Under depolarizing noise and faulty ballots the tally degrades (distortion and spoiled outcomes); no error correction or mitigation is applied here.} {The extended simulations of Section~\ref{subsec:extended} confirm the sampling bounds with confidence intervals and show quantitatively how memory, readout, and gate errors reduce the margin of victory; for example, memory errors accumulated over the voting stage limit the usable number of voters.}

\section{Discussion}
\label{sec:discussion}
The quantum voting protocol presented in this paper requires polynomial time number of basic quantum operations: Specifically, for each voter, the central voting system requires application of a multi controlled Z gate. For an $n$ qubit system, it is known that this type of multi controlled gates can be implemented by using $O(n)$ basic quantum operations \cite{nielsen2010quantum}. If we consider $N$ voters, the complexity for the voting part simply becomes $O(Nn)$.
At the end of the voting, the determination of the winner or tallying  is done by measuring or determining the tomography of a $\approx \log_2 K$-qubit state.
{While reading out the full tally requires an expected $2KN^2\ln(2K/\delta)$ protocol executions (including post-selection overhead; Section~\ref{subsubsec:exact_tallying}), the winner can be determined in an expected $\frac{2K}{\Delta^2}\ln\frac{K-1}{\delta}$ executions, which is independent of $N$ for a fixed margin $\Delta$ (Section~\ref{subsubsec:winner_only}). Both scalings are matched by lower bounds for any procedure based on the accepted outcomes (up to constants and the $\ln K$ term).}

{The $O(KN^2\log(K/\delta))$ executions required for exact vote counting become prohibitive for very large $N$, and this growth is intrinsic to sampling-based readout (Proposition~\ref{prop:exact_lower}). The winner-only determination alleviates this but still requires a non-negligible margin of victory. Therefore, it can be used efficiently when the votes are not close among the candidates.}

{Each protocol execution replays the voting stage from the stored ballots (Section~\ref{sec:models}), so the total gate count is the number of executions times $O(Nn)$. The executions are independent and can be distributed over several machines.}

{Abstentions (empty votes) are not analyzed separately. An empty ballot leaves the branch of its voter unchanged, so such a voter contributes nothing to the post-selected pool. With $N_c\le N$ cast ballots, the acceptance probability becomes $N_c/(NK)$, which reveals only the turnout, and the accepted outcomes follow the distribution $n_k/N_c$. The bounds of Section~\ref{sec:complexity} then apply with $N_c$ in place of $N$ in the vote fractions and an expected $KN/N_c$ executions per accepted outcome.}

{The anonymity results of Section~\ref{sec:security} hold under the stated trust assumptions: correct machines whose internal memory is trusted (Section~\ref{sec:models}) and a semi-honest Center. Several limitations remain. A malicious Center or machine could prepare non-uniform initial superpositions or perform intermediate measurements to correlate identity and vote information. Achieving security against such an adversary would require techniques from verifiable blind quantum computation~\cite{arapinis2021definitions} or fully decentralized multiparty protocols~\cite{centrone2022quantum}. {The status of each security property (correctness, fairness, ballot secrecy, eligibility, individual and universal verifiability, robustness, and tally integrity) in the unified adversarial model of Section~\ref{subsec:unified} is summarized in Table~\ref{tab:security}. In particular, individual verifiability is not provided, universal verifiability holds only for the decision derived from the public transcript, and fairness and tally integrity rest on the trusted machines.}}

Furthermore, the protocol does not prevent a coercer from forcing a voter to reveal their vote or to vote in a particular way. This is a known open problem in quantum voting.
More precisely, {our protocol does not claim to be receipt-free or coercion-resistant. {Voters have no access to the machine's quantum registers}, so no quantum receipt exists, but coercion resistance then depends on classical polling-place assumptions, for example that a voter cannot prove to a third party how the machine recorded the ballot. The phase-flip encoding provides no mechanism for a voter to plausibly deny or later change a coerced vote.} Coercion resistance and receipt-freeness remain open across essentially all quantum voting proposals, including recent self-tallying and GHZ-based schemes~\cite{wang2016self,centrone2022quantum}; formal definitions of these properties in the quantum setting are discussed in Ref.~\cite{arapinis2021definitions}. Classical mitigation techniques---such as supervised voting environments, revoting policies where only the last cast vote counts, or fake-credential mechanisms---could in principle be layered on top of our protocol, and we leave a rigorous integration as future work.

{In the distributed setting, the machines} must store their qubits coherently from distribution until tallying. This is another limitation since decoherence during this interval reduces the fidelity of the final state and may compromise both correctness and security. Active quantum error correction would be necessary for large‑scale deployment.

A variant in which individual voters hold qubits would remove the trust in the machines. It would, however, require the voters' participation in every run (the state cannot be replayed from stored records), open a voter-side attack surface (for example, overvoting by applying a phase to the control qubit, which adds weight to several candidates at once and produces no spoiled outcomes), and require a derivation showing that independent voter--center pairs realize the global state assumed in the tally. We leave this variant to future work.

Future work could explore hybrid quantum‑classical protocols that combine the anonymity guarantees of quantum superposition with the efficiency of classical tallying, as well as integration with post‑quantum cryptographic primitives for enhanced verifiability.

\subsection{Comparison with Existing Quantum Voting Protocols}
Table~\ref{tab:comparison} summarizes the main design dimensions of representative quantum voting protocols and highlights where the proposed scheme differs.

Note that the approach introduced in this paper differs from existing quantum voting protocols: As noted in the introduction, entanglement‑based protocols such as those in Refs.~\cite{vaccaro2007quantum,hillery2006towards} accumulate phases from multiple voters and require phase estimation to extract the tally. In contrast, this method uses a simpler gate‑based approach (controlled‑Z and Hadamard gates) to flip phases, making it potentially more amenable to near‑term quantum hardware. Moreover, our phase‑flip encoding provides a natural and efficient vote‑counting mechanism without complex multi‑particle measurements. 

In terms of resource scaling, the phase-accumulation approach of Ref.~\cite{vaccaro2007quantum} requires the precision of the final collective (phase-estimation-type) measurement to grow with the electorate size in order to resolve individual vote increments, whereas our protocol keeps every quantum operation elementary (Hadamard and controlled-$Z$) and shifts the cost entirely to classical repetitions of a shallow circuit, with the explicit sample-complexity bounds derived in Section~\ref{subsec:measurement_complexity}. The qudit-based scheme of Ref.~\cite{hillery2006towards} similarly requires $d$-level systems and collective operations that are challenging on current qubit hardware.

While GHZ‑based protocols \cite{centrone2022quantum} offer different security guarantees, this model can also be integrated with such entangled states. Finally, our protocol can be combined with existing quantum security primitives, such as quantum key distribution \cite{mehic2020quantum}, to further enhance security in distributed settings.

\paragraph{Quantitative comparison of resources and guarantees.}
Table~\ref{tab:quant} compares the resources and the security guarantees of representative protocols, using only what the cited works state; entries marked ``n.s.'' are not stated there. The comparison shows a trade-off rather than a uniform advantage. The protocols of Refs.~\cite{vaccaro2007quantum,hillery2006towards,horoshko2011quantum} use one entangled ballot state per election and a single collective tally measurement, but need entanglement across all $N$ voters (or bosonic modes, or qudits of dimension larger than $N$) and offer weaker or probabilistic security statements. The self-tallying protocol of Ref.~\cite{wang2016self} and the authority-free protocol of Ref.~\cite{centrone2022quantum} provide verifiability and explicit soundness, but need $O(N^2)$ particles or $N$-qubit GHZ states and many verification rounds. The present protocol needs only $\lceil\log_2N\rceil+1+\lceil\log_2K\rceil$ qubits per run in the centralized setting and shallow, elementary gates, but pays with $O(Nn)$ gates in every run, with $O(KN^2\log(K/\delta))$ runs for exact tallies or $O(K\Delta^{-2}\log(K/\delta))$ runs for the winner, and with trusted machines; it does not provide individual or universal verifiability (Table~\ref{tab:security}).

\begin{sidewaystable*}
\centering
\caption{{Quantitative comparison with representative protocols, as stated by the cited works (n.s.: not stated). $N$: voters, $K$: candidates, $n=\lceil\log_2N\rceil$. ``This work'' is evaluated for $N=100$, $K=3$, $\Delta=0.1$, $\delta=0.05$: $n=7$, $g(n)=12n+6=90$ two-qubit gates per ballot operation, and $2K\Delta^{-2}\ln\frac{K-1}{\delta}\approx2.2\times10^3$ runs for the winner or $2KN^2\ln\frac{2K}{\delta}\approx2.9\times10^5$ runs for the exact tally.}}
\label{tab:quant}
{%
\small
\resizebox{0.9\textheight}{!}{%
\begin{tabularx}{\textheight}{|>{\hsize=0.60\hsize\raggedright\arraybackslash}X|>{\hsize=1.10\hsize\raggedright\arraybackslash}X|>{\hsize=1.00\hsize\raggedright\arraybackslash}X|>{\hsize=1.10\hsize\raggedright\arraybackslash}X|>{\hsize=0.90\hsize\raggedright\arraybackslash}X|>{\hsize=1.30\hsize\raggedright\arraybackslash}X|}
\hline
\textbf{Protocol} & \textbf{Quantum resource} & \textbf{Operations; rounds} & \textbf{Tally measurement; repetitions} & \textbf{Trust} & \textbf{Security stated by the authors} \\
\hline
Vaccaro et al.~\cite{vaccaro2007quantum} & $N'\ge N$ bosons in 2 modes & 1 local phase shift per voter & 1 collective measurement; repetitions n.s. & Tallyman; honest voters (binary variant) & Maximally mixed reduced states; two-voter collusion detected with probability $1-\frac1{N'+1}$; one vote per voter not enforced; no verifiability analysis \\
\hline
Hillery et al.~\cite{hillery2006towards} & $N$ qudits, $d>N$ & 1 local phase operation per voter & Authority sets the phase; ``several'' repetitions & Curious authority; malicious authority left open & Reduced states $\propto\mathbb{I}$; partial protection against multiple voting; verifiability not fully achieved \\
\hline
Horoshko and Kilin~\cite{horoshko2011quantum} & $N$ qubits (Bell pairs for checks) & 1 send--tally--return--test sequence & 1 projective measurement & Curious tallyman & Tallyman measurements detected (miss probability $\frac12$ per checked qubit); no anonymity for a single voting act \\
\hline
Wang et al.~\cite{wang2016self} & $n^2(1+\delta_0+\delta_1)+n$ particles ($\delta_{0,1}$ unspecified) & Test rounds, then 1 broadcast per voter & Anyone tallies from the broadcast & Simultaneous broadcast; abort on failed test & Privacy, verifiability, fairness, self-tallying (mostly informal arguments) \\
\hline
Centrone et al.~\cite{centrone2022quantum} & $N$-qubit GHZ states; $\approx4000$ per vote ($N=4$) & $N$ voting and verification rounds; $\log_2K$ repetitions & Parities of outcomes & Untrusted source; no authority & Completeness, soundness, $\zeta$-privacy, verifiability, receipt-freeness \\
\hline
\textbf{This work (centralized)} & $n+1+\lceil\log_2K\rceil$ qubits per run ($10$ in the example) & $O(Nn)$ gates per run ($\approx9\times10^3$); depth grows with $N$ & Interference, then candidate readout; $2.2\times10^3$ (winner) or $2.9\times10^5$ (exact) runs & Trusted machine; semi-honest Center & Proved: anonymity (transcript plus auxiliary information), correctness (failure $\le\delta$), eligibility. Not provided: individual or universal verifiability, receipt-freeness \\
\hline
\textbf{This work (distributed)} & $N+n_t$ pairs per run ($n_t\approx5.7\times10^3$ test pairs) & As above, plus pair distribution for every run & As above & As above, plus authenticated channels and quantum memory & As above, plus CHSH completeness and soundness for tested pairs \\
\hline
\end{tabularx}}}
\end{sidewaystable*}

\section{Conclusion}

We have presented a comprehensive quantum voting protocol that leverages quantum superposition and entanglement to provide secure, anonymous voting in both centralized and distributed settings. In particular, the model includes a novel phase-flip based voting mechanism that enables efficient tallying through simple quantum measurements. It establishes information-theoretic security guarantees for voter anonymity {under the trusted-machine assumptions of Section~\ref{sec:models};} anonymity is formalized as vote-permutation indistinguishability of the tallying transcript (Theorem~\ref{thm:transcript}) and, with the identity-dephasing augmentation, the identity register is exactly maximally mixed (Theorem~\ref{thm:dephasing}).  The overall architecture is described for both single-machine and {distributed-machine} implementations. Therefore, it can be used along with existing quantum security frameworks such as quantum key distribution \cite{mehic2020quantum} and can be integrated with other entanglement based protocols such as \cite{centrone2022quantum}.
\section*{Conflict of interest statement}
The authors declare no conflict of interest.

\section*{Funding information}
This research received no specific grant from any funding agency in the public, commercial, or not-for-profit sectors.

\section*{Data availability statement}
The data that support the findings of this study are openly available in GitHub at \url{https://github.com/uuu4/quantum-voting-protocol}, archived on Zenodo (DOI: \href{https://doi.org/10.5281/zenodo.17373814}{10.5281/zenodo.17373814}). The repository contains the source code and Jupyter notebooks for the numerical simulations of the original submission. The scripts and result files of the extended simulations added in the revision (Section~9.5) are provided in the same repository, in the folder \texttt{extended\_simulations}. All results are simulations; no experimental data were generated.

\section*{CRediT author contribution statement}
\textbf{Ali Emre Aydin:} Software, Investigation, Data curation, Validation, Visualization, Writing -- original draft, Writing -- review \& editing.\\
\textbf{Ammar Daskin:} Conceptualization, Methodology, Formal analysis, Supervision, Project administration, Writing -- original draft, Writing -- review \& editing.

\section*{Acknowledgments}
AI-based tools were used to improve readability and to correct typographical mistakes. In the revision, a large language model (Claude, Anthropic) was additionally used as an assistant for literature and reference checking, for checking derivations and numerical constants, and for implementing and running the extended simulations of Section~9.5 under the authors' direction. The authors reviewed and verified all content and take full responsibility for it. 

\bibliographystyle{unsrtnat}
\bibliography{main}

\end{document}